\documentclass[a4paper,11pt]{article}
\pdfoutput=1 % if your are submitting a pdflatex (i.e. if you have
\usepackage{jcappub} % for details on the use of the package, please
\usepackage{tikz}
\usepackage{amsmath}
\usepackage{amssymb}

\usepackage[T1]{fontenc} % if needed
\usepackage{breqn}

\def\Ak{{\mathcal{A}_k}}
\def\Bk{{\mathcal{B}_k}}

\title{Steepest Growth in the Primordial Power Spectrum from Excited States at a Sudden Transition}

\author[a,b, c]{Mattia Cielo,}
\author[a,b]{Gianpiero Mangano,}
\author[a,b]{Ofelia Pisanti,}
\author[d]{David Wands}

\affiliation[a]{INFN - Sezione di Napoli, Complesso Univ. Monte S. Angelo, I-80126 Napoli, Italy}
\affiliation[b]{Dipartimento di Fisica ``Ettore Pancini”, Università degli studi di Napoli ``Federico II”, Complesso Univ. Monte S. Angelo, I-80126 Napoli, Italy}
\affiliation[c]{Instituto de Física Téorica UAM/CSIC, calle Nicolás Cabrera 13-15, Cantoblanco, 28049,
Madrid, Spain}
\affiliation[d]{Institute of Cosmology and Gravitation, Dennis Sciama Building, University of Portsmouth, PO1 3FX, United Kingdom}

\emailAdd{mattia.cielo@na.infn.it}

\abstract{Sudden phase transitions during inflation can give rise to strongly enhanced primordial density perturbations on scales much smaller than those directly probed by cosmic microwave background anisotropies. In this paper, we study the effect of the incoming quantum state on the steepest growth found in the primordial power spectrum using a simple model of an instantaneous transition during single-field inflation. We consider the case of a general de Sitter-invariant initial state for the inflaton field (the $\alpha$-vacuum), and also an incoming state perturbed by a preceding transition. For the $\alpha$-vacua we find that $k^6$ growth is possible for $\alpha>0$, while $k^4$ growth is seen for $\alpha\leq0$, including the standard case of an initial Bunch-Davies vacuum state. The features of an enhanced primordial power spectrum on small scales are thus sensitive to the initial quantum state during inflation. We calculate the scalar-induced gravitational wave power spectrum for each case.}

\begin{document}
\maketitle
\flushbottom

%%%%%%%%%%%%%%%%%%%%%%%%%%%%%%%%%%%%%%%%%%%%%%%%%%% COFFEE STAIN
%\coffeestainA{0.5}{0.8}{-25}{7cm}{8.3cm}

\section{Introduction}
What is the steepest growth of the primordial scalar power spectrum after an inflationary stage? This question has received considerable attention~\cite{Byrnes:2018txb, Carrilho:2019oqg, Ozsoy:2023ryl,Tasinato:2020vdk, Ballesteros:2021fsp, Cole:2022xqc, Zhai:2023azx, Palma:2020ejf} since the abundance and mass spectrum of primordial black holes (PBHs) and the associated stochastic gravitational wave (GW) background~\cite{LISACosmologyWorkingGroup:2023njw} are known to be sensitive to the shape of the primordial power spectrum as well as the height of its peak \cite{Garcia-Bellido:1996mdl, Green:2020jor, Escriva:2022duf, Ivanov:1994pa, Kinney:2005vj, Garcia-Bellido:2017aan, Motohashi:2017kbs, Karam:2022nym, Karam:2023haj}. A consistent interpretation of observational bounds requires a treatment that goes beyond the simplistic assumption of an almost monochromatic scalar power spectrum~\cite{Cole:2023wyx, Ozsoy:2023ryl, Escriva:2022duf, Zhai:2023azx}.
Previous work argued that rapid growth in the power spectrum is expected to be at most $\sim k^4$ \cite{Byrnes:2018txb,Cole:2022xqc}, while subsequently it was argued that $\sim k^5(\log k)^2$ was the steepest possible growth~\cite{Carrilho:2019oqg}.
%(or that other models from inflation which predict a different slope have a small impact on the final computation of the PBH abundance \cite{Cole:2022xqc}. )
Such strong scale-dependence often requires features in the inflaton potential or in the sound speed \footnote{For an alternative mechanism see Ref.~\cite{Ashoorioon:2018uey, Ashoorioon:2019xqc}}, leading to a sudden phase transition and the breakdown of slow roll, mixing adiabatic and isocurvature modes of the scalar field perturbations\cite{Jackson:2023obv}. Tasinato has shown~\cite{Tasinato:2020vdk} that multiple transitions during inflation can lead to much stronger enhancements in the power spectrum, indicating that the slope of the spectrum has a memory of the history of non-slow-roll phases during inflation. Sudden (non-adiabatic) phase transitions lead to particle production and a non-trivial incoming quantum state which can be amplified at subsequent transitions.
Motivated by these considerations and recent developments in the study of non-Bunch-Davies initial states of quantum perturbations ~\cite{Cielo:2022vmo, Cielo:2023enz, Holman:2007na, Yin:2023jlv, Ashoorioon:2014nta,  Kanno:2022mkx, Alberghi:2003am, Kundu:2011sg, Brandenberger:2012aj, Tanaka:2000jw, BouzariNezhad:2018zsi, Akama:2020jko, Shukla:2016bnu}, we will extend the analysis proposed in Ref.~\cite{Byrnes:2018txb} by incorporating the additional degeneracy inherent in quantum field theory on curved spacetime. 
It is well understood in fact that after the quantum-to-classical transition for the cosmological perturbations, the choice of the initial vacuum state is quite crucial~\cite{Lesgourgues:1996jc}. 
In particular, we study the so-called $\alpha$-vacua corresponding to de Sitter invariant initial states~\cite{Allen:1985ux}. 
This extension allows us to explore a broader range of scenarios and enrich our understanding of the quantum field dynamics.
We will study an idealised model of an instantaneous transition driven by a piecewise-linear potential, originally proposed by Starobinsky~\cite{Starobinsky:1992ts}, which provides a simple realisation of an inflationary model with a sudden transition from slow-roll to ultra-slow-roll, leading to a rapid growth in the primordial power spectrum. This model, despite its simplicity, can exhibit a rich phenomenology in the resulting primordial power spectrum for the scalar curvature perturbations and yield non-trivial results for both the production of the primordial black holes and the scalar-induced gravitational waves at second order. 
The paper is organized as follows. In section~\ref{sec:single} we introduce all the essential quantities needed to study the primordial scalar spectrum at the end of inflation, and give the general expressions for matching Bogoliubov coefficient across a sudden transition. We then introduce the Starobinsky piecewise-linear potential and emphasize the role of the choice of the incoming quantum state. We show analytically how the growth of the power spectrum can be estimated from the relation between the Bogoliubov coefficients before and after a sudden transition. We study the shape of the scalar power spectra resulting from different choices for the initial $\alpha$-vacua and the impact that this has on the stochastic GW background produced at second order.  We calculate the effective GW energy density, $\Omega_{GW}$, using the public code SIGWfast \cite{Witkowski:2022mtg}. In section~\ref{sec:multiple} we generalize Starobinsky's model of a single transition to a piecewise-linear potential including an arbitrary number of transitions and derive a general recurrence formula for obtaining the Bogoliubov coefficients after each step as functions of the previous ones. We then apply this analysis to the specific case of $n=2$ transitions. We conclude in section~\ref{sec:conclusions}.

\section{Single sudden transition during inflation}
\label{sec:single}
%\MCcomment{after many changes I opted for the convention using $\mathcal{A}$ for the bogoliubov coefficients and $\alpha$ for the alpha vacuum parameter and then $A_i$ for the slopes. I hope you also agree with me even if this choice differs from the recent literature.}
We work within the single-field paradigm of cosmic inflation. Specifically, we consider a minimally-coupled inflaton field governed by the Klein-Gordon equation in a Friedmann-Lemaitre-Robertson-Walker (FLRW) cosmology
\begin{equation}
    \ddot{\phi} + 3 H \dot{\phi} + \frac{dV}{d\phi} = 0
    \,,
    \label{eq. Klein-Gordon complete}
\end{equation}
where the Hubble rate $H\equiv\dot{a}/a$ is given by the Friedmann constraint
\begin{equation}
    H^2 = \frac{8\pi}{3} \left( V+\frac12 \dot\phi^2 \right) \,.
\end{equation}
The inflationary dynamics can be described in terms of two dimensionless slow-roll parameters
\begin{equation}
    \epsilon_1 \equiv -\frac{\dot{H}}{H^2} , \qquad \epsilon_2 \equiv \frac{\dot{\epsilon}}{\epsilon H} = -6 - \frac{2}{H \dot{\phi}} \frac{dV}{d\phi}
    \,,
    \label{def: slow-roll parameters}
\end{equation}
where we require $\epsilon_1<1$ for an accelerated expansion.
We will work in the quasi-de Sitter limit, in which $\epsilon_1\to0$ so $H$ is effectively constant, but we will allow $\epsilon_1$ to vary in time, hence $\epsilon_2$ is not necessarily small. It is small in the usual slow-roll approximation ($|\epsilon_2|\ll1$), but becomes large during a period of ultra-slow roll inflation ($\epsilon_2\approx-6$) \cite{Dimopoulos:2017ged}.
Each wave mode of the curvature perturbation, $\mathcal{R}$, with comoving wavenumber $k$, obeys the evolution equation
\begin{equation}
    \mathcal{R}''_k + 2 \frac{z'}{z} \mathcal{R}_k + k^2 \mathcal{R}_k = 0 \,,
    \label{eq. R original}
\end{equation}
where $z(\tau)\equiv a\dot\phi/H$
%$z(\tau) = a \sqrt{2 \epsilon}M_{Pl}$ 
and primes denote derivatives with respect to conformal time, $\tau=\int dt/a$. Note that in de Sitter ($\epsilon_1\to0$) we have $a=-1/H\tau$ and thus $\tau$ runs from $-\infty$ to $0$ as $a$ evolves from $0$ to $\infty$.
Through a suitable rescaling, $v = \mathcal{R}z$, we can cast (\ref{eq. R original}) into the Sasaki-Mukhanov mode equation
\begin{equation}
\label{modeequation}
    v''_k + \left(k^2 - \frac{z''}{z}\right) v_k = 0 \,.
\end{equation}
%This equation represents the mode functions, denoted as $v_k$, required for quantizing the perturbations.
This second-order differential equation is solved by a linear combination of two independent solutions, which can be written as
\begin{equation}
    \label{eq. general solution}
    v_k(\tau) =\frac{\Ak f_k (\tau) + \Bk f_k^* (\tau)}{\sqrt{2k}}  \,,
%    =\mathcal{A}_k H^{(1)}_{\nu} (x) + \mathcal{B}_k H^{(2)}_{\nu} (x) 
\end{equation}
%where we used the variable $x = - k\tau$. 
%So, we are left with the following quantum operator evolving in a quasi-de Sitter spacetime. 
%\begin{equation}
%    \hat{\mathcal{R}}_k = u_k \hat{a}_k + u^*_k \hat{a}^{\dag}_k
%\end{equation}
where the complex functions $f_k(\tau)$ are normalised such that the Wronskian obeys $f_k^*f_k'-f_k^{*'}f_k=2ik$. In particular we will require that $f_k(\tau)\to e^{-ik\eta}$ in the short-wavelength limit $(z''/z)/k^2\to0$.
The commutation relations for the corresponding quantum field operator, $\hat{v}$, and its canonical momentum then require
\begin{equation}
\label{normAB}
    |\Ak^2|-|\Bk^2| = 1 \,.
\end{equation}
Finally, we must specify the initial state of the field, which determines the behaviour of the quantum operator, $\hat{v}$, on the ground state of the theory, $|0 \rangle$. The most common choice is the Bunch-Davies vacuum ($\Ak=1$, $\Bk=0$) which corresponds to the minimum energy state~\cite{Mukhanov:2007zz}, but for now, we will leave the initial state arbitrary.
We will consider models where a sudden transition occurs due to a potential, $V(\phi)$, which is continuous but with a sudden change in the derivative at $\phi=\phi_1$, ~\cite{Starobinsky:1992ts, Pi:2022zxs, Domenech:2023dxx}. We are not interested in effects coming from the detailed form of the potential around $\phi=\phi_1$, so we will treat the transition as instantaneous. We expect this to be a good description of wavemodes $k\ll a/\Delta t$ where $\Delta t$ is the duration of the transition in terms of proper time, and we define a sudden transition to be one for which $\Delta t\ll H^{-1}$. 
We thus have the junction conditions at $\phi_1$: 
\begin{equation}
    [V]^+_- = 0 \qquad \left[\frac{dV}{d\phi}\right]^+_- = \Delta V'
\end{equation}
This leaves the field and its time derivative continuous, $[\phi]_-^+=[\dot\phi]_-^+=0$, but the second time
derivative is discontinuous, $[\ddot\phi]_-^+=-\Delta V'/3H$. Hence $z(\tau)$ is continuous, but $z'(\tau)$ is discontinuous:
\begin{equation}
\left[ z \right]^+_- = 0, \qquad \left[\frac{z'}{z} \right]^+_- = - \frac{a_1 \Delta V'}{\dot{\phi}_1} \equiv \Delta_1
    \label{eq: mode_eq}
\end{equation}
Note that the scalar field density and pressure are continuous across the transition, but the
time-derivative of the pressure is discontinuous, hence we will refer to this as a second-order phase transition
(the second time-derivative of the energy density is discontinuous).
From integrating the mode equation (\ref{eq: mode_eq}) across the transition (or from requiring the continuity
of the comoving curvature perturbation, $\mathcal{R}$, and its derivative) we find:
\begin{equation}
\label{mode_junction}
    [v_k]^+_- = 0, \qquad [v'_k]^+_- = - \Delta_1 v_k
\end{equation}
Without loss of generality, we will require that the particular solution of the Sasaki-Mukhanov mode equation \eqref{modeequation}, $f_k$, and its derivative, $f'_k$, are continuous across the transition.
In terms of the general solution (\ref{eq. general solution}) we thus have
%\DWcomment{I have swapped the +/- ordering in the next two equations to match my notes, and the preceding equations. The results that then follow match my notes, so I believe it is just the next two equations that had +/- the wrong way round.}
\begin{equation}
\left[\mathcal{A}_k f_k + \mathcal{B}_k f^*_k\right]_{-}^+ = 0 , \label{eq:vkjunction}
\end{equation}
\begin{equation}
\left[ \mathcal{A}_k f'_k + \mathcal{B}_k f^{'*}_k \right]_-^+ = -\Delta_1 (\mathcal{A}_k f_{k1} + \mathcal{B}_k f^*_{k1}) \,. \label{eq:vkprimejunction}
\end{equation}
where $f_{k1}=f_k(\tau_1)$.
Following \cite{Starobinsky:1992ts} we will assume that $\dot\phi<0$ and the scalar field evolves from $\phi>\phi_1$ to $\phi<\phi_1$. Before the transition, for $\phi>\phi_1$, we write
\begin{equation}
\mathcal{A}_{k^+} = \mathcal{A}_k \,, \quad \mathcal{B}_{k^+} = \mathcal{B}_k \,,\label{eq:1.6}
\end{equation}
while after the transition, for $\phi<\phi_1$, we set 
\begin{equation}
\mathcal{A}_{k^-} = \tilde{\mathcal{A}}_k \,, \quad \mathcal{B}_{k^-} = \tilde{\mathcal{B}}_k \,. \label{eq:1.7}
\end{equation}
From Eqs.~\eqref{eq:vkjunction} and~\eqref{eq:vkprimejunction} we can establish a general relationship between the Bogoliubov coefficients before and after the transition\footnote{It is straightforward to check that the normalisation condition \eqref{normAB} still holds after the transition, as it should for a quantum field, and we have $|\tilde{\mathcal{A}}_k^2|-|\tilde{\mathcal{B}}_k^2| = 1$.}
%\begin{eqnarray}
%\tilde{\mathcal{A}}_k &=& \mathcal{A}_k - \frac{i\Delta_1 f^*_{k1}}{2k}(f_{k1} \mathcal{A}_k + f^*_{k1} \mathcal{B}_k) \,, \nonumber
%\\
%\tilde{\mathcal{B}}_k &=& \mathcal{B}_k + \frac{i\Delta_1 f_{k1}}{2k}(f_{k1} \mathcal{A}_k + f^*_{k1} \mathcal{B}_k) \,. \label{A+Bafter}
%\end{eqnarray}
%\DWcomment{or [anticipating the matrix notation used later]}
\begin{equation}
\label{A+Baftermatrix}
   \begin{bmatrix} \tilde{\mathcal{A}}_k \\ \tilde{\mathcal{B}}_k \end{bmatrix}
 =
 \begin{bmatrix} 
 1-i\Delta_1 f^*_{k1}f_{k1}(2k)^{-1} & -i\Delta_1 f^*_{k1}f^*_{k1}(2k)^{-1} \\ 
 +i\Delta_1 f_{k1}f_{k1}(2k)^{-1} & 1+i\Delta_1 f^*_{k1}f_{k1}(2k)^{-1} \end{bmatrix}
 \begin{bmatrix} 
 {\mathcal{A}}_k \\ {\mathcal{B}}_k \end{bmatrix}
\,.
\end{equation}

\subsection{Starobinsky piecewise-linear potential}
\label{subsec:Starobinsky}
The Starobinsky piecewise-linear potential serves as a convenient illustrative example of a sudden (non-adiabatic) transition~\cite{Starobinsky:1992ts}. This model is characterized by the potential:
\begin{equation}
V(\phi) = 
\begin{cases}
V_0 + A_1 (\phi - \phi_1) , \qquad \phi > \phi_1\\
V_0 + A_2 (\phi - \phi_1) , \qquad  \phi < \phi_1
\end{cases}
\label{potential}
\end{equation}
In our scenario, $\dot\phi<0$ and $A_1=A_+>0$ represents the slope of the potential before the transition and $A_2=A_-$ represents the slope after the transition at $\phi=\phi_1$. 
%We emphasize that this model serves as a basic representation of a phase transition between two cosmological scenarios, portrayed as an instantaneous process. A more nuanced and realistic depiction of the transition will be explored in future research endeavors.
We assume that during the initial regime, $\phi>\phi_1$, our field follows a slow-roll evolution. In this regime, we can neglect the acceleration term in the Klein-Gordon equation~(\ref{eq. Klein-Gordon complete}), leading to the friction being equal and opposite to the potential slope
\begin{equation}
\label{eq:SRattractor1}
    \dot\phi\simeq-\frac{A_1}{3H} \quad {\rm for}\ \phi>\phi_1 \,. 
\end{equation} 
Since we will assume that we are close to de Sitter throughout, we require $A_1\ll M_{Pl}H^2$ and thus during slow roll both $\epsilon_1\ll1$ and $|\epsilon_2|\ll1$.
%
%Upon reaching the transition point between the two linear regimes, the gradient of the potential suddenly decreases. 
When the field reaches $\phi=\phi_1$ there is a sudden change in the potential gradient, $\Delta V'=A_1-A_2$.
After the transition ($t>t_1$) the Klein-Gordon equation \eqref{eq. Klein-Gordon complete} can be integrated to give the simple solution: 
%\DWcomment{I have given the simple solution for $\dot\phi$ after a single transition case here, prior to giving the multiple-transition case later:}
\begin{equation}
\dot{\phi} = - \left( \frac{A_1-A_2}{3H} \right) e^{- 3H (t - t_1)} - \frac{A_2}{3H} \quad {\rm for}\ \phi<\phi_1 \,.
\end{equation}
For $0<A_2\ll A_1$ the potential gradient becomes much smaller than the friction term immediately after the transition, and the slow-roll approximation is no longer applicable. In this phase the field's deceleration is driven by the friction term, $\ddot\phi\approx-3H\dot\phi$, violating slow roll and, from Eq.~\eqref{def: slow-roll parameters}, leading to $\epsilon_2 \approx -6$, while $\epsilon_1$ remains small. This is commonly referred to as Ultra-Slow Roll inflation (USR) \cite{Dimopoulos:2017ged}. 
However, this USR phase is transient, as the time-derivative of the scalar field decays rapidly, $\dot\phi\propto e^{-3Ht}$, and hence the field's acceleration also decays rapidly, and at late times the system approaches a new slow-roll attractor with $3H\dot\phi\simeq-A_2$ and $|\epsilon_2| \ll1$. 
%Let's delve into the dynamics of curvature perturbations within this specific scenario. In the forthcoming analysis, we'll examine the repercussions of selecting the initial quantum state and then narrow our focus to the particular potential under consideration.
%The evolution of primordial quantum perturbations manifests in two distinct regimes, each influencing the solution of the Mukhanov-Sasaki equation \cite{Mukhanov:2007zz}. 
To compute the primordial power spectrum, we solve the Mukhanov-Sasaki mode equation \eqref{modeequation} before and after the transition. 
For the piece-wise linear potential \eqref{potential}, corresponding to a massless inflaton field in the de Sitter limit ($\epsilon_1\to0$), the time-dependent mass in the mode equation \eqref{modeequation}, $z''/z=2/\tau^2$, remains invariant~\cite{Wands:1998yp,Leach:2001zf} before and after the transition at $\phi=\phi_1$. Thus the general solution to the mode equation is given by the expression in Eq.~\eqref{eq. general solution} where in both regimes
\begin{equation}
\label{freefk}
f_k(\tau) = \left( 1 - \frac{i}{k\tau} \right) e^{-ik\tau} \,.
\end{equation}
We see that on small scales and at early times ($k\tau\to-\infty$)  $f_k\to e^{-ik\tau}$, while on large scales and late times ($k\tau\to0$) $f_k=-f_k^*\to -i/k\tau$. Thus the dimensionless power spectrum for the primordial curvature perturbation is
\begin{equation}
\label{def:PS}
    {\cal P}_{\mathcal{R}}(k) = \frac{4\pi k^3|v_k^2|}{(2\pi)^3z^2} \,,
    \end{equation}
Before the transition, in slow roll, the initial conditions for the modes are set by the choice of the vacuum state, e.g., the Bunch-Davies vacuum, where ${\mathcal{A}}_k=1$ and ${\mathcal{B}}_k=0$, and we recover the scale-invariant power spectrum ${\cal P}_{\mathcal{R}}=( {H^2}/{2\pi\dot\phi})^2$.
However at late times, after the transition, the power spectrum is given by~\cite{Maggiore:2018sht}
\begin{equation}
\label{PRlate}
    {\cal P}_{\mathcal{R}}(k) \to \left( \frac{H^2}{2\pi\dot\phi} \right)^2 |\tilde{\mathcal{A}}_k - \tilde{\mathcal{B}}_k|^2 \quad {\rm as}\ k\tau\to 0 \,.
\end{equation}
%and the final primordial power spectrum can be computed by evaluating the modulus of the difference between the two Bogoliubov coefficients
%\begin{equation}
%{\cal P}_{\mathcal{R}}(k)  \propto |\tilde{\mathcal{A}}_k - \tilde{\mathcal{B}}_k|^2 ,
%\end{equation}
The matrix \eqref{A+Baftermatrix} matching the Bogoliubov coefficients across the sudden transition, using the mode function \eqref{freefk} for a free field in de Sitter, becomes
\begin{equation}
\label{A+Baftermatrix-linear}
   \begin{bmatrix} \tilde{\mathcal{A}}_k \\ \tilde{\mathcal{B}}_k \end{bmatrix}
 =
 \begin{bmatrix} 
 1-(i\Delta_1/2k_1) \left(1+\kappa^2\right)/\kappa^3 \quad & (i\Delta_1/2k_1) (1+i\kappa)^2 e^{-2i\kappa}/\kappa^3 \\ 
 -(i\Delta_1/2k_1) (1-i\kappa)^2 e^{2i\kappa} / \kappa^3 \quad & 1+(i\Delta_1/2k_1) \left(1+\kappa^2\right) /\kappa^3 \end{bmatrix}
 \begin{bmatrix} 
 {\mathcal{A}}_k \\ {\mathcal{B}}_k \end{bmatrix}
\,,
\end{equation}
where $k_1$ represents the wavenumber crossing outside the Hubble scale at the time of the transition, $k_1=a_1H$, and here and in the following equations we have introduced the normalized wavenumber $\kappa \equiv k/k_1$.
The sudden transition in the acceleration of the field results in a discontinuity in the time-derivative of $z(\tau)$ given by Eq.~\eqref{eq: mode_eq}. For the piecewise linear potential \eqref{potential} we have
\begin{equation}
\label{sr_delta}
\Delta_1 = \frac{3k_1(A_1 - A_2)}{A_1} \,.
\end{equation}
After the transition the Bogoliubov coefficients are thus set by Eqs.~\eqref{A+Baftermatrix-linear}, where using \eqref{sr_delta} we obtain
\begin{eqnarray}
\label{A+Bafter}
    \tilde{\mathcal{A}}_k &=& {\mathcal{A}}_k - \frac{3i}{2\kappa^3} \left( \frac{A_1-A_2}{A_1} \right) \left\{ (1+\kappa^2) {\mathcal{A}}_k - (1+i\kappa)^2e^{-2i\kappa} {\mathcal{B}}_k \right\} \,,\nonumber\\
    \tilde{\mathcal{B}}_k &=& {\mathcal{B}}_k + \frac{3i}{2\kappa^3} \left( \frac{A_1-A_2}{A_1} \right) \left\{ (1+\kappa^2) {\mathcal{B}}_k - (1-i\kappa)^2e^{2i\kappa} {\mathcal{A}}_k \right\} \,.
\end{eqnarray}
The final power spectrum \eqref{PRlate} for an arbitrary initial state is determined by the combination
%
%\begin{equation}
%\tilde{\mathcal{A}}_k - \tilde{\mathcal{B}}_k = \mathcal{A}_k - \mathcal{B}_k + \frac{\Delta_1}{2ik}(f_{k1} + f^*_{k1}) \{f_{k1} \mathcal{A}_k + f^*_{k1} \mathcal{B}_k\} \,.
%\label{eq. difference A - B}
%\end{equation}
%where $f_{k1}\equiv f_k(\tau_1)$.
%
%It's evident from this framework that following the transition, the curvature perturbation can be expressed as a linear combination of mode functions, each weighted by corresponding Bogoliubov coefficients. This observation proves valuable, particularly considering that 
%
%\begin{equation}
%\label{A-Bafter}
%\tilde{\mathcal{A}}_k - \tilde{\mathcal{B}}_k = \mathcal{A}_k - \mathcal{B}_k - \frac{3i}{2} \frac{A_1 - A_2}{A_1} \left( \frac{1 + \frac{k^2}{k_1^{2}}}{1 + \frac{k^2}{k_1^{2}} + \frac{i}{k_1 k^2} e^{2ik/k_1}} \mathcal{A}_k + \frac{1 + \frac{k^2}{k_1^{2}}}{1 - \frac{i}{k_1 k^2} e^{-2ik/k_1}} \mathcal{B}_k \right) 
%\,. 
%\end{equation}
%
%\begin{equation}
%\label{A-Bafter}
%\begin{split}
%\tilde{\mathcal{A}}_k - \tilde{\mathcal{B}}_k = \mathcal{A}_k - \mathcal{B}_k - \frac{3i}{2\kappa} \frac{A_1 - A_2}{A_1} \Big[ \Big(1 + \frac{1}{\kappa^2} + \Big(1 +  \frac{i}{\kappa} \Big)^2e^{2i\kappa} \Big)\mathcal{A}_k & + \\  \Big(1 + \frac{1}{\kappa^2} + \Big(1 -  \frac{i}{\kappa} \Big)^2e^{-2i \kappa} \Big)\mathcal{B}_k \Big]  \,.
%\end{split}
%\end{equation}
%
\begin{equation}
\label{A-Bafter}
%\label{A-Bafter-Bessel}
    \tilde{\mathcal{A}}_k - \tilde{\mathcal{B}}_k = \left( \mathcal{A}_k - \mathcal{B}_k \right) \left\{ 1+3\left(\frac{A_1 - A_2}{A_1}\right)\kappa j_1(\kappa)y_1(\kappa) \right\}
    -i \left( \mathcal{A}_k + \mathcal{B}_k \right) \left\{ 3\left(\frac{A_1 - A_2}{A_1}\right)\kappa j_1^2(\kappa) \right\}
    \,,
\end{equation}
where
\begin{equation}
\label{def:j1y1}
    j_1(\kappa)=\frac{\sin \kappa}{\kappa^2}-\frac{\cos \kappa}{\kappa} \,,
    \qquad
    y_1(\kappa)=-\frac{\cos\kappa}{\kappa^2}-\frac{\sin\kappa}{\kappa} \,,
\end{equation}
are spherical Bessel functions of the first and second kind.
For super-Hubble modes at the transition ($\kappa<1$) we can Taylor expand \eqref{def:j1y1}, which substituted into Eq.~\eqref{A-Bafter} gives
\begin{equation}
\begin{split}
\tilde{\mathcal{A}}_k - \tilde{\mathcal{B}}_k 
%&= \frac{A^-}{A^+} - \frac{A^+ - A^-}{A^+} \frac{2}{5} \frac{k^2}{k_*^2} \left(1 + O\left(\frac{k^2}{k_*^2}\right)\right) (\mathcal{A}_k - \mathcal{B}_k) \\
%&\quad - \frac{A^+ - A^-}{A^+} \frac{i}{3} \frac{k^3}{k_*^3} \left(1 + O\left(\frac{k^2}{k_*^2}\right)\right) (\mathcal{A}_k + \mathcal{B}_k) .
&= \left\{ \frac{A_2}{A_1} - \frac{2}{5} \left( \frac{A_1 - A_2}{A_1} \right) \kappa^2 \left(1 + {\cal O}\left(\kappa^2 \right)\right) \right\}\left(\mathcal{A}_k - \mathcal{B}_k\right) \\
&\quad - \frac{i}{3} \left( \frac{A_1 - A_2}{A_1}\right) \kappa^3 \left(1 + {\cal O}\left(\kappa^2 \right)\right) \left(\mathcal{A}_k + \mathcal{B}_k\right) \,.
\end{split}
\label{eq:superHubbleA-B}
\end{equation}
We see that $\tilde{\mathcal{A}}_k - \tilde{\mathcal{B}}_k\to (A_2/A_1)(\mathcal{A}_k - \mathcal{B}_k)$ as $\kappa\to0$ and the primordial power spectrum \eqref{PRlate} remains unchanged in the very large-scale limit\footnote{Recall that before the transition $\dot\phi=-A_1/3H$ while $\dot\phi\to-A_2/3H$ as $\tau\to0$ after the transition}. However, for $A_1\gg A_2$ there will be a rapid rise in the primordial power spectrum on a range of scales larger than the Hubble scale at the transition, $A_2/A_1<\kappa^2<1$, with the slope depending on the incoming state, given by $\mathcal{A}_k$ and $\mathcal{B}_k$. 
%Now, we proceed to apply the concepts elucidated thus far to the case of the Starobinsky potential. With the general solutions for the curvature perturbation $\mathcal{R}_k$ in both phases in hand, we determine the final Bogoliubov coefficients by enforcing matching conditions at the transition point. This entails: 
%\begin{equation}
%    \mathcal{R}^{(1)}_k(\tau_1) = \mathcal{R}^{(2)}_k(\tau_1) \qquad , \qquad  {\mathcal{R} '}^{(1)}_k(\tau_1) = {\mathcal{R}'}^{(2)}_k(\tau_1)
  %  \label{eq. matching}
%\end{equation}

\subsubsection{Bunch-Davies initial state}
\begin{figure}
    \centering
    \includegraphics[scale=0.5]{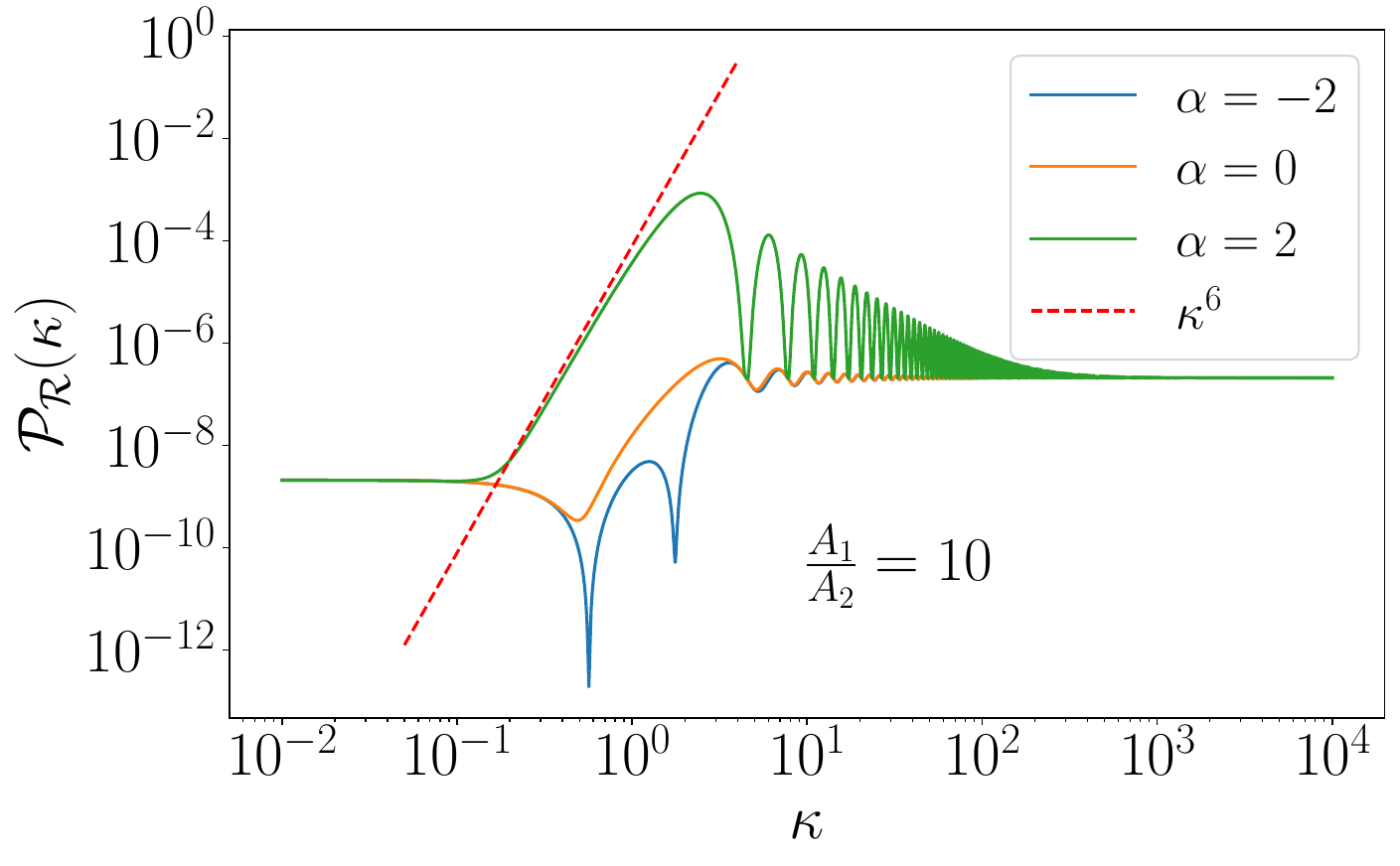}
    \caption{The power spectrum of the curvature perturbation $\mathcal{P}_{\mathcal{R}}$ at the end of inflation, Eq.~\eqref{def:PS}, as a function of the normalized wavenumber, $\kappa = k/k_1$, where $k_1$ is the wavenumber which crosses the horizon at the sudden transition.
    The plot shows power spectra obtained for different initial $\alpha$-vacuum states \eqref{alpha-vacuum}, varying $\alpha$ for $\beta=0$. 
    %The top (green) curve shows $\alpha=+2$, the middle (yellow) curve shows $\alpha=0$ (the Bunch-Davies vacuum) and the bottom (blue) curve shows $\alpha=-2$. 
    In this plot, we fix the ratio between the slopes $A_1/A_2 = 10$ and we have fixed $H^3e^{\alpha}/A_1$ such that $\mathcal{P}_{\mathcal{R}}\to2\times 10^{-9}$ for $\kappa\to0$.} 
    \label{fig: PS_1}
\end{figure}

Having obtained the general solution to the mode functions~\eqref{eq. general solution} before and after the transition, and thus the primordial power spectrum~\eqref{def:PS}, we must now specify the initial state for the quantum field at early times.
The simplest choice for the initial state is the one adopted by Bunch and Davies which is equivalent to selecting the minimum energy state for the field~\cite{Mukhanov:2007zz}.
In this case the Bogoliubov coefficients~\eqref{eq. general solution}, before the transition, take the simple form: 
\begin{equation}
\label{def:BDvacuum}
    \mathcal{A}_k = 1 \qquad \mathcal{B}_k = 0 
\end{equation}
After the transition, the quantum state is determined by the junction conditions~\eqref{mode_junction} leading to the new Bogoliubov coefficients \eqref{A+Bafter}~\cite{Starobinsky:1992ts}
\begin{eqnarray}
   \tilde{\mathcal{A}}_k &=&  1 - \frac{3i (A_1-A_2)}{2 A_1} \frac{\left( 1 + \kappa^2 \right)}{\kappa^3}
   \label{eq. Ak BD single}
   \,, \\
    \tilde{\mathcal{B}}_k &=& -\frac{3i(A_1 - A_2)}{2A_1} \frac{(1 - i\kappa)^2e^{2i\kappa}}{\kappa^3}
    \,.
    \label{eq. Bk BD single}
\end{eqnarray}
The sudden transition~\eqref{eq: mode_eq} results in a state that mixes negative and positive frequency modes. %
It is apparent that even when starting from the Bunch-Davies vacuum state, a sudden transition ($A_2\neq A_1$) induces excitations in the modes corresponding to particle production ($\tilde{\mathcal{B}}_k\neq0$) in the inflaton field on sub-horizon scales after the transition, and non-adiabatic perturbations on super-Hubble scales~\cite{Jackson:2023obv}. 
Substituting Eqs~\eqref{def:BDvacuum} into \eqref{A-Bafter} to compute the primordial scalar power spectrum at the end of inflation \eqref{PRlate}
%\begin{equation}
%    \begin{aligned}
%        \mathcal{P}_{\mathcal{R}} &= \frac{9 H^6}{16 \pi^2 \text{A}_1^2 \text{A}_2^2 \kappa^6} \times \left[            \left(3 \left(\kappa^2-1\right) (\text{A}_1-\text{A}_2) \sin (2 \kappa) + 6 \kappa (\text{A}_1-\text{A}_2) \cos (2 \kappa) + 2 \text{A}_1 \kappa^3 \right)^2 \right. \\         &\left. + 36 (\text{A}_1-\text{A}_2)^2 (\sin (\kappa) - \kappa \cos (\kappa))^4 \right]
%    \end{aligned}
%    \label{PS without alpha}
%\end{equation}
we obtain
\begin{equation}
    \mathcal{P}_{\mathcal{R}} = \frac{9 H^6}{4 \pi^2 \text{A}_1^2 \text{A}_2^2} \left\{ \left[ A_1 +3(A_1-A_2) \kappa j_1(\kappa) y_1(\kappa)\right]^2 + \left[3(A_1-A_2)\kappa j_1^2(\kappa)\right]^2 \right\} \,,
        \label{PS without alpha}
\end{equation}
where 
\begin{equation}
    \kappa j_1(\kappa)y_1(\kappa) = \frac{(\kappa^2-1)\sin(2\kappa) + 2\kappa\cos(2\kappa)}{2\kappa^3}
     \,,
    \qquad
    \kappa j_1^2(\kappa) =
    \frac{(\kappa\cos\kappa- \sin\kappa)^2}{\kappa^3} \,,
\end{equation}
and we have used the late-time solution for the time-derivative of the field after the transition, $\dot{\phi} \to -A_2/(3H)$.
The power spectrum \eqref{PS without alpha} is shown by the middle (orange) curve in Figure~\ref{fig: PS_1}.
From the Taylor expansion on super-Hubble scales at the transition, Eq.~\eqref{eq:superHubbleA-B}, we see that if modes start in the Bunch-Davies vacuum state \eqref{def:BDvacuum}, then on very large scales, $\kappa^2\ll A_2/A_1\ll1$, the spectrum remains scale-invariant. However for smaller scales, but still larger than the Hubble scale at the transition, $\kappa^2<1$, the real part of  $\tilde{\mathcal{A}}_k - \tilde{\mathcal{B}}_k$ in Eq.~\eqref{eq:superHubbleA-B} vanishes for $\kappa^2\simeq(5/2)A_2/A_1$, corresponding to a dip in the power spectrum~\cite{Starobinsky:1992ts,Leach:2001zf,Pi:2022zxs}. This is followed by a rise in $|\tilde{\mathcal{A}}_k - \tilde{\mathcal{B}}_k|\propto\kappa^2$, hence a rise in $\mathcal{P}_{\mathcal{R}}\propto\kappa^4$, for smaller scales, but still larger than the Hubble scale~\cite{Leach:2001zf} 
\begin{equation}
    \mathcal{P}_{\mathcal{R}} = \left( \frac{3H^3}{5\pi A_2} \right)^2 \kappa^4
%    \left(\frac{k}{k_1} \right)^2 
    \,, \quad {\rm for}\ \  \frac{A_2}{A_1} \ll \kappa^2 \ll 1 \,.
\end{equation}
The power spectrum rises to a peak at $\kappa\sim\pi$~\cite{Starobinsky:1992ts,Pi:2022zxs} and then exhibits damped oscillations about the asymptotic value on sub-Hubble scales
\begin{equation}
    \mathcal{P}_{\mathcal{R}} \to \frac{9 H^6}{4 \pi^2 \text{A}_2^2} 
    \,, \quad {\rm for}\ \  \kappa \to \infty \,.
\end{equation}
Finally, we note from Eq.~\eqref{eq. Bk BD single} that the particle number density after the transition $k^3|\tilde{\mathcal{B}}_k|^2\propto\kappa$ for $\kappa\gg1$. The corresponding energy density is thus formally divergent as $\kappa\to\infty$. This is a result of modelling the transition as instantaneous, which excites arbitrarily high energy modes. In practice the particle production will be suppressed on length scales much smaller than the duration of the transition, $\kappa\gg a/(k_1\Delta t)$.

\subsubsection{$\alpha$-vacuum inital state}
We will now extend our calculation of the primordial power spectrum to explore the additional degeneracy induced by the choice of the initial state of the quantum field \eqref{eq. general solution}. As previously remarked, the commutation relation for the field and its canonical momentum place the constraint \eqref{normAB} on the Bogoliubov coefficients.
For this reason, we adopt the parametrization in terms of hyperbolic functions~\cite{Allen:1985ux}
\begin{equation}
\label{alpha-vacuum}
    \mathcal{A}_k = \cosh(\alpha) \qquad \mathcal{B}_k = e^{i\beta} \sinh (\alpha)
\end{equation}
Here, $\alpha$ is a real free parameter, characterizing the initial state within a family of degenerate $\alpha$-vacuum states obeying the normalisation condition \eqref{normAB}.
This degeneracy arises due to the absence of temporal Killing vectors in the FLRW foliation of the spacetime~\cite{Danielsson:2002kx, Broy:2016zik}.
Here we will choose $\alpha$ and $\beta$ to be scale-invariant (independent of the wavenumber, $k$) which corresponds to a de Sitter invariant initial state~\cite{Allen:1985ux}.
The Bunch-Davies vacuum \eqref{def:BDvacuum} represents the particular choice $\alpha=0$ for the initial state.
Although we might be tempted to choose any value for the parameters $\alpha$ and $\beta$, we must exercise caution to ensure that we do not disrupt the inflationary phase by introducing an excessive particle density. 
The particle energy density is given by~\cite{Lyth:1998xn, Chung:1999ve}:
\begin{equation}
\label{eq:particle density}
%    n_{\delta \phi} = \frac{1}{\pi^2 a^3} \int^{\Lambda/a}_0 d k \, k^2 |\mathcal{B}_k|^2 
    \varepsilon_{\delta \phi} \sim \frac{1}{a^4} \int^{a\Lambda}_0 d k \, k^3 |\mathcal{B}_k|^2 
\end{equation}
where we take a physical UV cut-off, $\Lambda$.
%Since our intent is to prevent the overproduction of inflationary particles 
We need to guarantee that during the inflationary stage, the particle energy density is much less than that of the potential, $V_0$.
Hence we require $\sinh^2(\alpha)\ll V_0/\Lambda^4$. In practice, we will limit our discussion to the range $-2\leq \alpha \leq 2$.
Using the $\alpha$-vacuum initial state~\eqref{alpha-vacuum} before the transition, we can again compute the outgoing Bogoliubov coefficients \eqref{A+Bafter} after having applied the matching condition \eqref{mode_junction}
\begin{eqnarray}
\label{A after alpha}
    \tilde{\mathcal{A}}_k &=& \left[ 1 - \frac{3i}{2\kappa^3} \left( \frac{A_1-A_2}{A_1} \right) (1+\kappa^2) \right] \cosh(\alpha) 
    \nonumber \\
     && \quad + \frac{3i}{2\kappa^3} \left( \frac{A_1-A_2}{A_1} \right)(1+i\kappa)^2e^{-2i\kappa} e^{i\beta} \sinh(\alpha)   \,,\\
     \label{B after alpha}
    \tilde{\mathcal{B}}_k &=& \left[ 1  + \frac{3i}{2\kappa^3} \left( \frac{A_1-A_2}{A_1} \right) (1+\kappa^2) \right] e^{i\beta} \sinh(\alpha) \nonumber \\
     && \quad 
    - \frac{3i}{2\kappa^3} \left( \frac{A_1-A_2}{A_1} \right) (1-i\kappa)^2e^{2i\kappa} \cosh(\alpha) \,.
\end{eqnarray}
\begin{figure}
    \centering
    \includegraphics[scale=0.5]{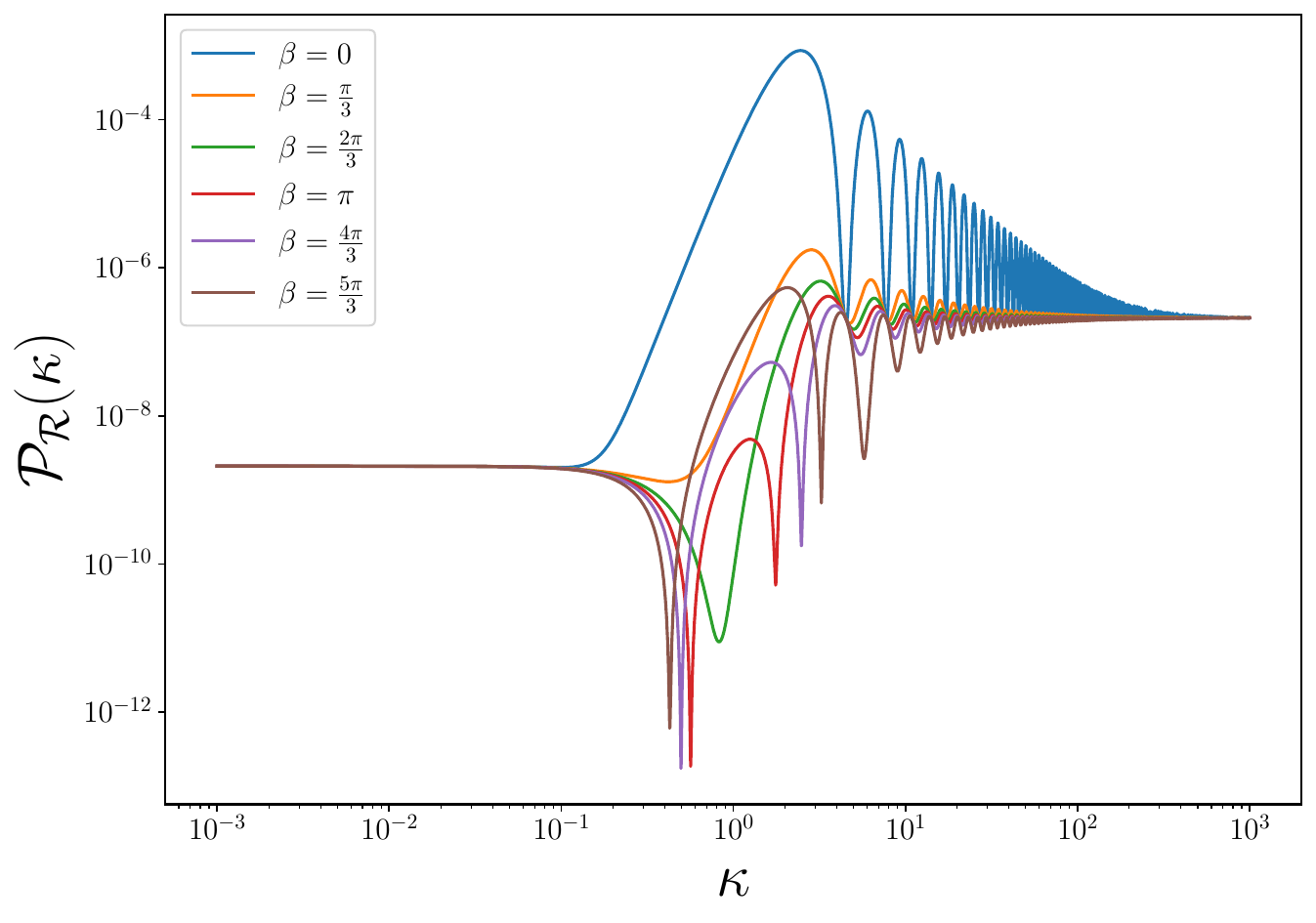}
    \caption{The power spectrum of the curvature perturbation $\mathcal{P}_{\mathcal{R}}$ \eqref{def:PS} as a function of the normalized wavenumber $\kappa = k/k_1$, where $k_1$ is the wavenumber which crosses the horizon at the sudden transition. 
    The plot shows different power spectra obtained by varying the $\beta$ parameter in the initial $\alpha$-vacuum states \eqref{alpha-vacuum}, while fixing the value of $\alpha = 2$. 
    In this plot, we fix the ratio between the slopes $A_1/A_2 = 10$ and we have fixed $H^3e^{\alpha}/A_1$ such that $\mathcal{P}_{\mathcal{R}}\to2\times 10^{-9}$ for $\kappa\to0$.} 
    \label{fig: PS_beta}
\end{figure}
Substituting \eqref{alpha-vacuum} into \eqref{A-Bafter} we obtain the power spectrum at the end of inflation~\eqref{PRlate}
\begin{eqnarray}
\label{PRlate-alphavacuum}
    \mathcal{P}_{\mathcal{R}} &=& \frac{9 H^6}{4 \pi^2 \text{A}_1^2 \text{A}_2^2} \Big\{ \Big[ \left(\cosh\alpha-\cos\beta\sinh\alpha\right) \left( A_1 +3(A_1-A_2) \kappa j_1(\kappa) y_1(\kappa)\right) 
    \nonumber  \\
    && \qquad \qquad \qquad \qquad  + 3\sin\beta\sinh\alpha (A_1-A_2)\kappa j_1^2(\kappa) \Big]^2 
    \nonumber \\
    && \qquad + \Big[ \sin\beta\sinh\alpha \left( A_1 +3(A_1-A_2) \kappa j_1(\kappa) y_1(\kappa)\right) \nonumber \\
    && \qquad \qquad \qquad \qquad + 3\left(\cosh\alpha+\cos\beta\sinh\alpha\right) (A_1-A_2)\kappa j_1^2(\kappa)\Big]^2 \Big\} \,.
\end{eqnarray}
Let us first consider the case where we fix the phase $\beta=0$. 
In this case Eq.~\eqref{PRlate-alphavacuum} reduces to
%\begin{equation}
%\begin{split}
%    \mathcal{P}_{\mathcal{R}} &=  \frac{1}{16 \pi^2 A_1^2 A_2^2 \kappa^6} \Big( 9 H^6 \left((\cosh(\alpha) - \sinh(\alpha))^2 \left(3 (\kappa^2 - 1) (A_1 - A_2) \sin(2\kappa) \right. \right. + \\
%    & \left. \left. + 6\kappa (A_1 - A_2) \cos(2\kappa) + 2 A_1 \kappa^3\right)^2 + 36 (A_1 - A_2)^2 (\sinh(\alpha) + \cosh(\alpha))^2 (\sin(\kappa) - \kappa \cos(\kappa))^4 \right)
%\end{split}
%\end{equation}
%
\begin{eqnarray}
    \label{PS with alpha}
    \mathcal{P}_{\mathcal{R}} &=& \frac{9 H^6}{4 \pi^2 \text{A}_1^2 \text{A}_2^2} \Big\{ e^{-2\alpha} \Big[  A_1 +3(A_1-A_2) \kappa j_1(\kappa) y_1(\kappa)
     \Big]^2 
%    \nonumber \\
%    && \qquad \qquad \qquad \qquad 
    + e^{2\alpha} \Big[ 3(A_1-A_2)\kappa j_1^2(\kappa)\Big]^2 \Big\} \,, \quad \quad
\end{eqnarray}
This expression for the primordial power spectrum differs from the result for a Bunch-Davies initial state, Eq.~(\ref{PS without alpha}), solely due to the presence of the two exponential functions of $\alpha$ in Eq.~(\ref{PS with alpha}). By setting $\alpha$ to zero, we recover the standard solution from the Bunch-Davies vacuum. 
Figure~\ref{fig: PS_1} illustrates how the selection of the $\alpha$ parameter can significantly influence the shape of the primordial spectrum \eqref{PS with alpha} after a sudden transition.
A \textit{negative} $\alpha$ parameter, leads to a deeper initial dip at $\kappa_2\simeq(5/2)A_2/A_1$, where the real part of  $\tilde{\mathcal{A}}_k - \tilde{\mathcal{B}}_k$ in Eq.~\eqref{eq:superHubbleA-B} vanishes. There is then an additional dip before the power spectrum rises to reach a peak very similar to the standard Bunch-Davies scenario, displaying similar oscillations with the same periodicity for $\kappa\gg1$. Conversely, the \textit{positive} $\alpha$ case has no dip and gives a steeper growth reaching a significantly higher peak compared with the one obtained from the Bunch-Davies vacuum.
We can understand this behaviour by writing the Taylor expansion on super-Hubble scales \eqref{eq:superHubbleA-B} for the case of a strong transition, $A_2\ll A_1$ to give
\begin{equation}
\tilde{\mathcal{A}}_k - \tilde{\mathcal{B}}_k 
= \left( \frac{A_2}{A_1} - \frac{2}{5} \kappa^2 + {\cal O}\left(\kappa^4 \right)\right) e^{-\alpha} 
%\\ &\quad 
-  \frac{i}{3} \left( \kappa^3 + {\cal O}\left(\kappa^5 \right)\right) e^\alpha 
\quad {\rm for}\ \kappa\ll 1 
\,.
\label{eq:superHubble-alpha}
\end{equation}
We see that if we start in an $\alpha$-vacuum with $e^\alpha\gg1$ (such that $|\mathcal{A}_k + \mathcal{B}_k| \gg |\mathcal{A}_k - \mathcal{B}_k|$), then the leading correction for $(A_2/A_1)e^{-2\alpha}\ll \kappa^3\ll1$ becomes $k^3$, giving a $k^6$ rise in the power spectrum \eqref{PS with alpha} on super-Hubble scales at the transition, rather than the $k^4$ rise seen for the Bunch-Davies vacuum. 
Because the $\kappa^3$ term in \eqref{eq:superHubbleA-B} is out of phase with respect to the $\kappa^0$ term, there is no longer a dip in the power spectrum between the plateau and the rise in the power spectrum
for $e^\alpha\gg1$, as seen in in Figure~\ref{fig: PS_1} for for the case $\alpha=2$.
Conversely for $e^\alpha\ll1$ (negative $\alpha$)
the dip at $\kappa^2\simeq(5/2)A_2/A_1$, which is already seen in the case of the initial Bunch-Davies vacuum state, becomes deeper since the imaginary term in Eq.~\eqref{eq:superHubble-alpha} which remains non-zero is suppressed.
The peak of the power spectrum in Figure~\ref{fig: PS_1} is enhanced for $\alpha>0$ since the amplitude of the scale-invariant spectrum as $\kappa\to0$ 
is suppressed relative to Bunch-Davies vacuum case ($\alpha=0$)
\begin{equation}
    \mathcal{P}_{\mathcal{R}} \to \frac{9 e^{-2\alpha}H^6}{4 \pi^2 \text{A}_1^2} 
    \,, \quad {\rm for}\ \  \kappa \to 0 \,,
\end{equation}
\begin{equation}
    \mathcal{P}_{\mathcal{R}} \to \frac{9 e^{-2\alpha}H^6}{4 \pi^2 \text{A}_2^2} 
    \,, \quad {\rm for}\ \  \kappa \to \infty \,.
\end{equation}
Finally, the late-time power spectrum \eqref{PRlate-alphavacuum} is shown for $\alpha=2$ and different values of $\beta$ in Figure~\ref{fig: PS_beta}. 
For $\beta\neq0$ the rise of the power spectrum and the peak is less pronounced than for $\beta=0$. For $\beta\approx\pi$ there is again a dip in the power spectrum where the real part of  $\tilde{\mathcal{A}}_k - \tilde{\mathcal{B}}_k$ in Eq.~\eqref{eq:superHubbleA-B} vanishes at $\kappa^2\simeq(5/2)A_2/A_1$ before the rise and the peak is suppressed, however we see that the dip moves to smaller values of $\kappa$ for $\alpha\sin\beta<0$ (or larger values of $\kappa$ for $\alpha\sin\beta>0$).
In the rest of this paper we will focus on two specific values for the phase, corresponding to $\beta=0$ and $\beta=\pi$. Given the degeneracy between the sign of $\alpha$ and the phase of $\beta$ we will fix $\beta=0$ but consider both positive and negative values for $\alpha$.

\subsubsection{Scalar-induced gravitational waves}
\begin{figure}
    \centering
    \includegraphics[scale=0.5]{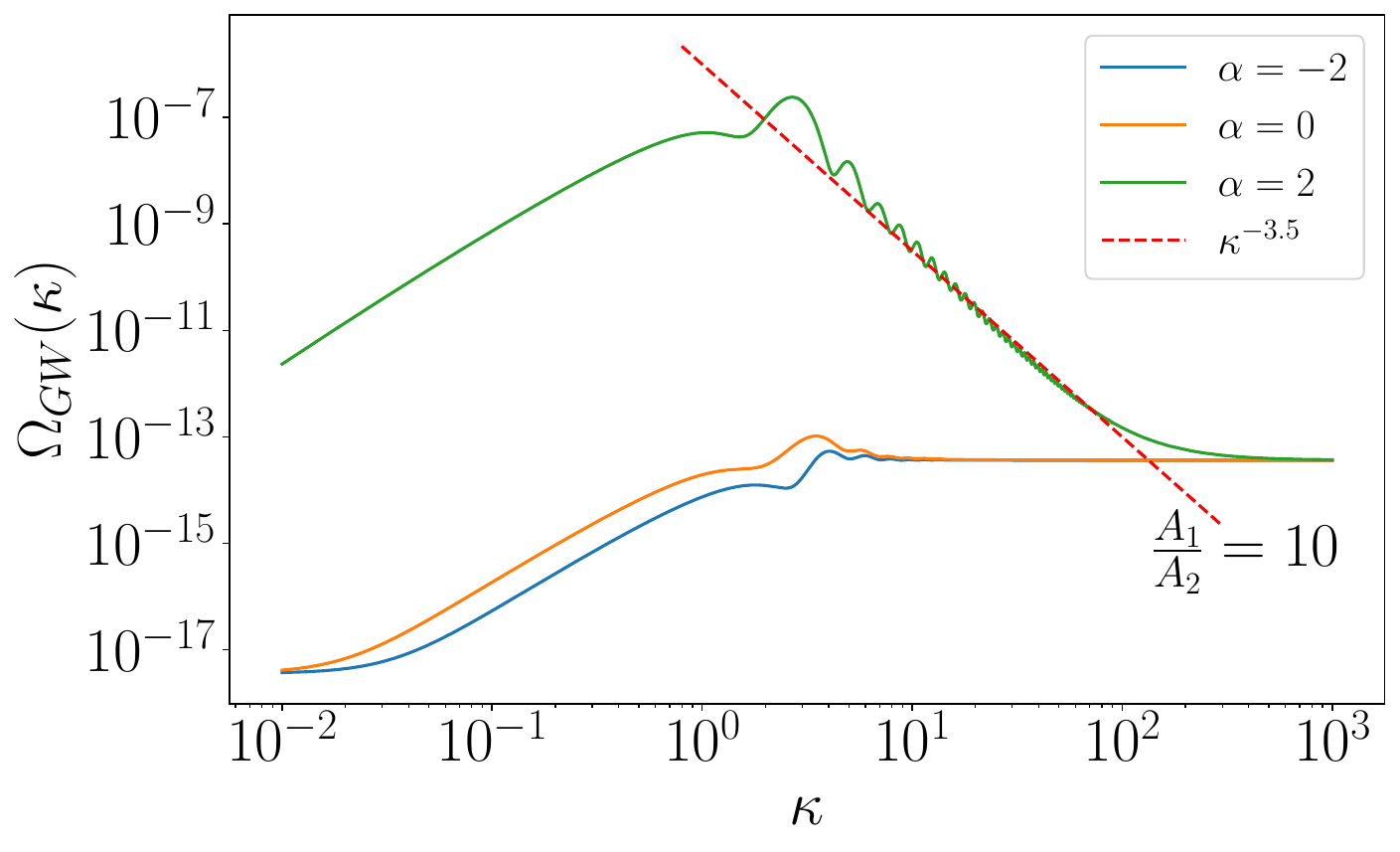}
    \caption{The gravitational waves energy density for the scalar induced gravitational waves $\Omega_{GW}$ as a function of the normalized wavenumber $\kappa = k/k_1$, where $k_1$ is the wavenumber which crosses the horizon at the sudden transition. 
    The plot shows the different GW power spectra obtained from the scalar power spectra shown in Figure~\ref{fig: PS_1}.}
    \label{fig: Omega1}
\end{figure}

While strong scale-dependent enhancements of the primordial scalar power spectrum have not been observed on the large, cosmological scales directly probed by the cosmic microwave background anisotropies, for example, they could be present on much smaller scales where they might be detectable through a scalar-induced stochastic gravitational wave background.
At second-order in perturbation theory, the gravitational wave (GW) power spectrum depends quadratically on the first-order scalar power spectrum~\cite{Tomita:1967wkp,Matarrese:1992rp,Matarrese:1993zf,Matarrese:1997ay}, so a scale-dependent feature in the scalar sector can also induce an enhancement and scale-dependence in the tensors. These second-order tensor perturbations could be much larger than those arising at first-order from the free fluctuations of the metric tensor~\cite{Ananda:2006af, Baumann:2007zm}. 
The dimensionless density of the induced GW background, $\Omega_{GW}$, is given by \cite{Ananda:2006af, Baumann:2007zm, Domenech:2021ztg, Fumagalli:2021mpc}:
\begin{equation}
\label{OmegaGWeq}
\Omega_{GW,eq}(k) = 3 \int_{0}^{\infty} \int_{1+v}^{|1-v|} \frac{T(u,v)}{u^2v^2} \mathcal{P}_{\mathcal{R}}(vk)\mathcal{P}_{\mathcal{R}}(uk) \, du \, dv
\end{equation}
where the kernel inside the integral is given by
\begin{equation}
\begin{split}
T(u,v) &= \frac{1}{4} \frac{1}{4v^2 - (1+v^2 - u^2)^2} \frac{4uv}{u^2 + v^2 - 3} \\
&\quad \times \left[ \ln \frac{3 - (u+v)^2}{3 - (u-v)^2} - \frac{4uv}{u^2 + v^2 - 3} + \pi^2 \Theta \left( u + v - \sqrt{3} \right) \right] 
\end{split} 
\,.
\end{equation}
%
%Our objective is to evaluate the impact of an additional parameter on the Scalar Induced Gravitational Waves (SIGWs). 
We have used the publicly available code SIGWfast \cite{Witkowski:2022mtg} to evaluate the GW density generated following a sudden transition in the Starobinsky piecewise linear model \eqref{potential} for an initial $\alpha$-vacuum state \eqref{alpha-vacuum}. In the following numerical examples, we consider three different values for the parameter $\alpha \in \{-2, 0, 2\}$ while fixing $\beta=0$, as in the previous sub-section. As shown in Figure~\ref{fig: Omega1}, we observe similar behaviour for $\alpha = -2$ and $\alpha = 0$, with only minor variations in amplitude, consistent with previous findings for the BD vacuum case~\cite{Pi:2022zxs}. However, the positive-$\alpha$ scenario shows a strong enhancement over $\alpha\leq0$ at its peak and a subsequent decline $\propto\kappa^{-3.5}$ with superimposed oscillations, before eventually converging to the same plateau as seen for $\alpha\leq0$, but only at much smaller scales. This is a significant difference with respect to $\alpha\leq0$, where the GW density shows only a small decrease for $\kappa\gg1$ from its peak near $\kappa\sim1$. 

\section{Generalization for multiple transitions} 
\label{sec:multiple}
We have seen that the scalar power spectrum after a sudden transition can be significantly enhanced when considering an excited incoming state with $\alpha>0$ in Eq.~\eqref{alpha-vacuum}. Thus it is interesting to explore the effect of multiple transitions during inflation, each of which will generate an excited state and can amplify any incoming excitations~~\cite{Tasinato:2020vdk}. 

\subsection{$n$ transitions}

In this section, we extend the Starobinsky model to encompass an arbitrary number of sudden transitions. As in the previous section~\ref{subsec:Starobinsky}, this generalized potential will be piece-wise linear:
\begin{equation}
\label{multipiecewisepotential}
V(\phi) = 
\begin{cases}
V_1 + A_1 (\phi - \phi_1) , \qquad \phi_1 < \phi\\
V_2 + A_2 (\phi - \phi_2) , \qquad \phi_2 < \phi < \phi_1 \\
... \\
V_n + A_{n+1} (\phi-\phi_n) , \qquad \phi < \phi_n\\
\end{cases}
\,,
\end{equation}
where $V_{i+1}=V_i+ A_{i+1} (\phi_{i+1} - \phi_i)$, so that the potential is continuous at $\phi=\phi_i$, but with a discontinuous first derivative if $A_{i+1}\neq A_i$.
We can adopt a methodology similar to that outlined in \cite{Pi:2022zxs} to compute $\dot{\phi}$ in each stage recursively.
We will work again in the quasi-de Sitter limit where $\epsilon_1\to0$ and we treat $H$ as effectively constant.
The Klein-Gordon equation~\eqref{eq. Klein-Gordon complete} can then be rewritten for $\phi_{i+1}<\phi<\phi_i$ as 
\begin{equation}
\frac{d}{dt} \left(\dot{\phi} + \frac{A_{i+1}}{3H}\right) + 3H \left(\dot{\phi} + \frac{A_{i+1}}{3H} \right) = 0 \,.
\label{new KG}
\end{equation}
%This transformation converts our Klein-Gordon equation into a differential equation for the variable $\dot{\phi} + \frac{A_2}{3H_0}$. 
This has the simple solution
\begin{equation}
\label{dotphii+1}
\dot{\phi} = - \frac{A_{i+1}}{3H} + \left( {\dot\phi}_i + \frac{A_{i+1}}{3H} \right) e^{- 3H (t - t_i)}
\end{equation}
where, $t_i$ is the cosmic time at the $i$-th transition which occurs at $\phi=\phi_i$ with velocity $\dot\phi={\dot\phi}_i$. 
Although we previously assumed that field was in the slow-roll attractor, ${\dot\phi}=-A_1/3H$, before a single transition at $\phi=\phi_1$, this will not necessarily be the case at a second or subsequent transition at $\phi=\phi_{i+1}$ for $i\geq1$ due to transient component, proportional to $e^{-3Ht}$, induced at the previous transition when $\dot\phi_i\neq A_{i+1}/3H$. 
Applying the simple solution \eqref{dotphii+1} multiple times, we see that after $i\geq2$ transitions the field's velocity for $\phi_{i+1}<\phi<\phi_i$ is given by
%\begin{equation}
%    \dot{\phi} = - \frac{A_{n+1}}{3 H} + e^{-3H(t - t_n)} \Big( \frac{A_{n+1} - A_n}{3 H} + \sum^{n-1}_{i = 1} \frac{A_{i+1} - A_i}{3 H}  \prod^{n-1}_{j=i} e^{-3H (t_{j+1} -t_j)} \Big)
%\end{equation}
\begin{equation}
    \dot{\phi} = - \frac{A_{i+1}}{3H} + \sum^{i}_{j = 2} \frac{A_{j+1} - A_j}{3 H} e^{-3H (t -t_j)} + \left( {\dot\phi}_1 + \frac{A_2}{3H} \right) e^{-3H (t -t_1)} \,.
\end{equation}
%\begin{equation}
%    \dot{\phi} = - \frac{A_{i+1}}{3H} \left( \frac{a^3-a_{j-1}^3}{a^3} \right)  - \sum_{j=2}^i \frac{A_j}{3H} \left( \frac{a_j^3-a_{j-1}^3}{a^3} \right) +\dot\phi_1 \left( \frac{a_1^3}{a^3} \right) \,.
%\end{equation}
If we assume that the field is in the slow-roll attractor for $\phi>\phi_1$, i.e., before the first transition, then this reduces to
%\begin{equation}
%    \dot{\phi} = - \frac{A_{i+1}}{3H} \left( \frac{a^3-a_{j-1}^3}{a^3} \right)  - \sum_{j=1}^i \frac{A_j}{3H} \left( \frac{a_j^3-a_{j-1}^3}{a^3} \right) \,.
%\end{equation}
%where we have set $a_0=0$, or better
\begin{equation}
\label{dotphi-sum}
    \dot{\phi} = - \frac{A_{i+1}}{3H} + \sum_{j=1}^i \frac{A_{j+1}-A_{j}}{3H} e^{-3H (t -t_j)}  \,.
\end{equation}
%
%Our objective is to present an analytical framework for recursively determining the form of the Bogoliubov coefficient for any number of transitions in the potential. Consequently, we undertake the task of solving the equation for the curvature perturbation $\mathcal{R}_k$ at each stage.
%Since the mode equation for $v_k$ remains the same across all stages, our primary focus lies in identifying the functions pertaining to the background, specifically, the velocity function of the field with respect to time and the function $z(\tau)$.
%Initially, we adopt a methodology akin to the one outlined in \cite{Pi:2022zxs} to compute the shape of $\dot{\phi}$ recursively.
%Let's outline the scheme briefly. We will employ the principle of induction to derive a general solution.
%During the initial phase, we can assume to be within the slow-roll attractor regime. Thus, the equation of motion takes the form:
%\begin{equation}
%\label{SRvelocity}
%3H \dot{\phi} = - V_{\phi} = - A_1
%\end{equation}
%The evolution of the field velocity is solely determined by the slope of the potential in the first stage, disregarding the contribution of the acceleration term.
%Following a transition, we can no longer neglect the acceleration term in the Klein-Gordon equation (entering the Ultra-Slow-Roll regime), necessitating consideration of the full Equation \ref{eq. Klein-Gordon complete}. In this regime, the equation also depends on the slope of the potential in the second stage:
%
We can now write down an expression for $z(\tau)\equiv a\dot\phi/H$ after $i\geq2$ transitions:
%\begin{equation}
%z(\tau) = \frac{a_0 \tau_0}{3H \tau} \left( -A_{n+1} + \left(\frac{\tau}{\tau_n}\right)^3 \left((A_{n+1} - A_n) + \sum_{i=1}^{n-1} \left((A_{i+1} - A_i) \prod_{j=i}^{n-1} \left(\frac{\tau_{j+1}}{\tau_j}\right)^3\right)\right)\right) \,.
%\label{multiple z}
%\end{equation}
%
\begin{equation}
    z(\tau) = \frac{A_{i+1}}{3H^3\tau} - \left( \sum_{j=1}^i \frac{A_{j+1}-A_{j}}{3H^3\tau_j^3} \right) \tau^2 \,.
\label{multiple z}
\end{equation}
%We can also express this equation in terms of the wavenumber $k_i$ of the mode which exits the horizon at conformal time $\tau_i = -1/k_i$. 
%
Within each piece-wise linear phase, $\phi_{i+1}<\phi<\phi_i$, the time-dependent mass in the mode equation \eqref{modeequation} remains invariant~\cite{Wands:1998yp,Leach:2001zf}, $z''/z=2/\tau^2$, but it diverges at the transitions when $\phi=\phi_i$ if $A_{i+1}\neq A_i$. The mode functions are thus given by the general solution \eqref{eq. general solution} where $f_k(\tau)$ in each phase is given by \eqref{freefk}, but at each transition we must match the Bogoliubov coefficients,
%to those in the preceding phase according to Eqs.~\eqref{eq:vkjunction} and \eqref{eq:vkprimejunction}, 
forming a chain of dependencies.
%
%Equation~\eqref{multiple z} shows that the impact of the difference between the slopes of neighbouring pieces of the potential is weighted by the ratio of the times (or equivalently the scales) at which the transitions happen. It is clear that the features of these models are encoded in the following parameter set: 
%\begin{equation}
%    \Big(A_{i+1} - A_i , \frac{\tau_{j+1}}{\tau_j} = \frac{k_j}{k_{j+1}} , \alpha \Big)
%\end{equation}
%At this point, we have all the ingredients to find the Bogoliubov coefficients after a sudden transition at $\phi=\phi_i$ as a function of the Bogoliubov coefficients after the preceding transitions at $\phi>\phi_{i-1}$. 
The generalisation of the expression for the Bogoliubov coefficients after a single transition, Eq.~\eqref{A+Baftermatrix}, to each step in a series of multiple transitions for an arbitrary potential is
\begin{equation}
\label{A+Brecursive}
   \begin{bmatrix} {\mathcal{A}}_{k,i+1} \\ {\mathcal{B}}_{k,i+1} \end{bmatrix}
 =
 \begin{bmatrix} 
 1-i\Delta_i f^*_{k,i}f_{k,i}(2k)^{-1} & -i\Delta_i f^*_{k,i}f^*_{k,i}(2k)^{-1} \\ 
 +i\Delta_i f_{k,i}f_{k,i}(2k)^{-1} & 1+i\Delta_i f^*_{k,i}f_{k,i}(2k)^{-1} \end{bmatrix}
 \begin{bmatrix} 
 {\mathcal{A}}_{k,i} \\ {\mathcal{B}}_{k,i} \end{bmatrix}
\,,
\end{equation}
where the generalisation of \eqref{eq: mode_eq} at $\phi=\phi_i$ gives
\begin{equation}
   \left[ z \right]^{i_+}_{i_-} = 0, \qquad \left[\frac{z'}{z} \right]^{i_+}_{i_-}
    = - \frac{a_i \Delta V_i'}{\dot{\phi}_i}
%    = \frac{a_i (A_{i+1}-A_i)}{\dot{\phi}_i} 
    \equiv \Delta_i \,.
\end{equation}
Specialising again to the case of the piecewise linear potential \eqref{multipiecewisepotential}, for a single transition where the field is initially in the slow-roll attractor approaching the transition at $\phi=\phi_1$ then we have $\dot{\phi}_1=-A_1/3H$ and for $i=1$ we recover \eqref{sr_delta}.
More generally, where $z(\tau)$ after $i\geq1$ transitions is given by Eq.~\eqref{multiple z}, we have
%\begin{equation}
%    \Delta_n = 3k_n \left( \frac{A_{n+1}-A_n}{A_n} \right) \left\{ 1+ \left( \frac{3H\dot\phi_{n-1}-A_n}{A_n} \right) \left( \frac{k_{n-1}}{k_n} \right)^3 \right\}^{-1}     \,.
%\end{equation}
\begin{equation}
    \Delta_i = 3k_i \left( A_{i}-A_{i+1} \right) \left\{ {A_i} + \sum_{j=1}^{i-1}\left( A_{j}-A_{j+1} \right) \left( \frac{k_j}{k_i} \right)^3 \right\}^{-1} 
    \,.
\end{equation}

\subsection{Two transitions}
\begin{figure}
    \centering
    \begin{minipage}[b]{0.49\textwidth}
        \centering
        \includegraphics[width=\textwidth]{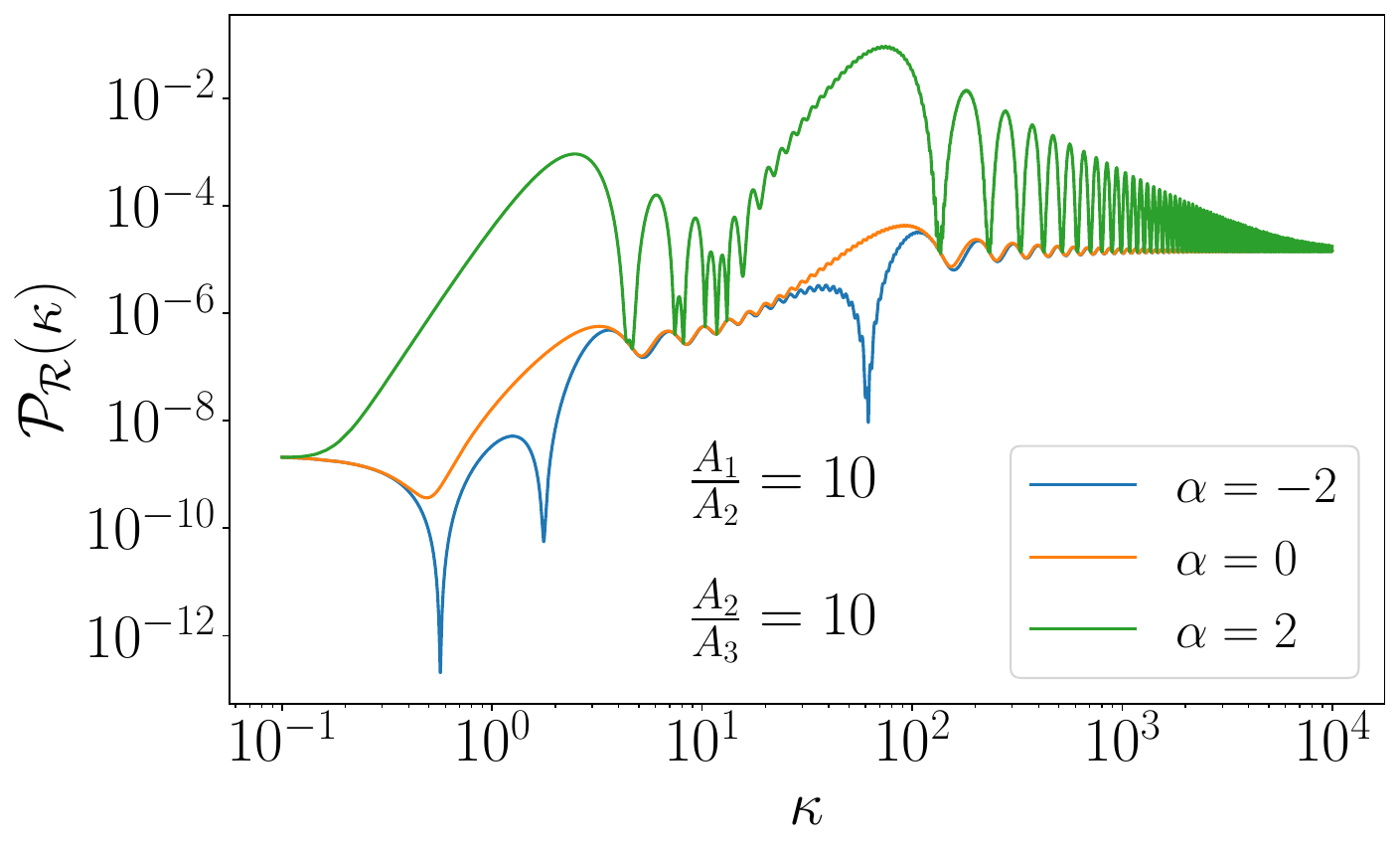}
    \end{minipage}
    \begin{minipage}[b]{0.49\textwidth}
        \centering
        \includegraphics[width=\textwidth]{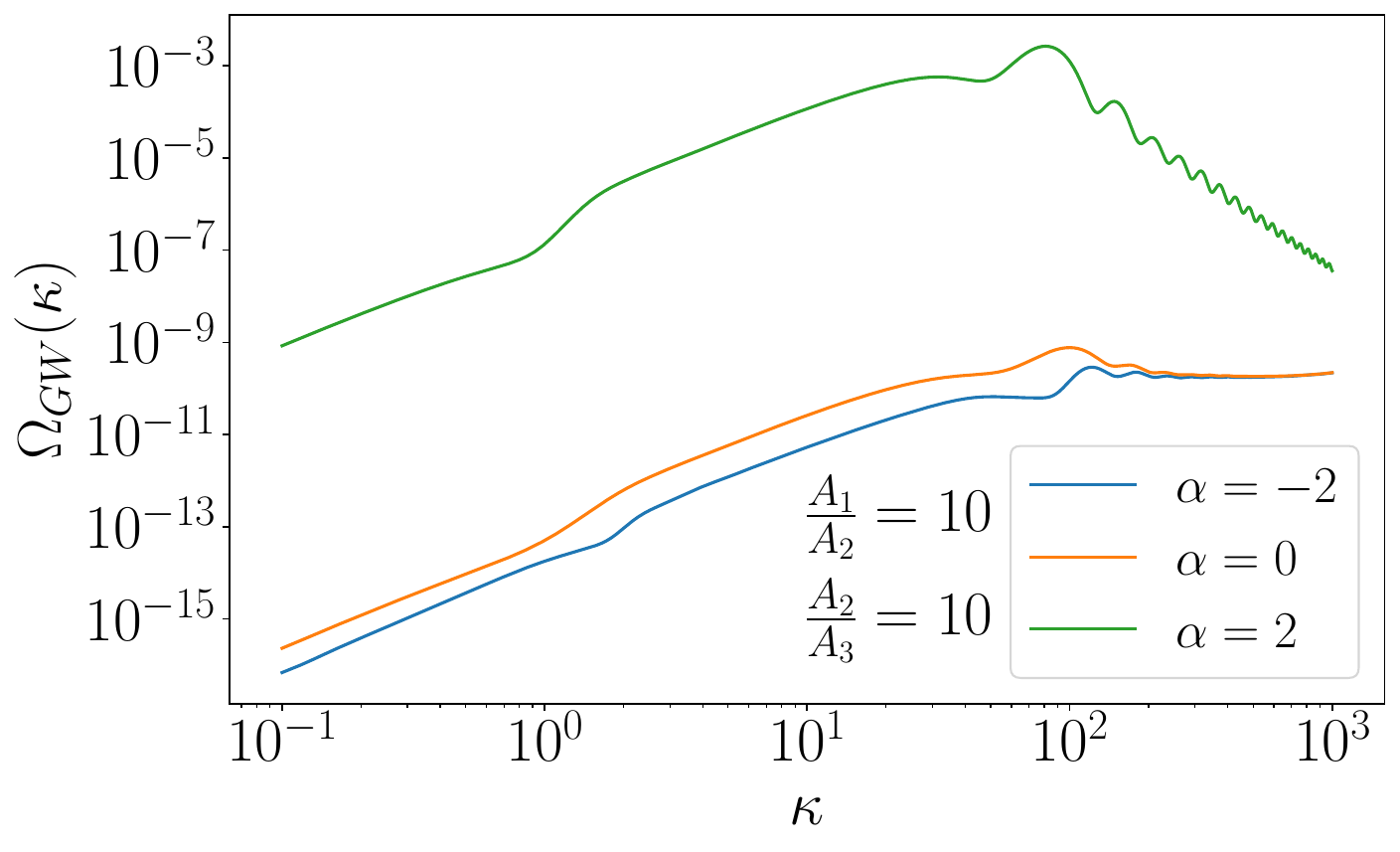}
    \end{minipage}
    \caption{\textit{Left}: The primordial power spectrum, $\mathcal{P}_{\mathcal{R}}$. \textit{Right}: The dimensionless energy density for the scalar induced gravitational waves, $\Omega_{GW}$. 
    Both spectra are shown as a function of the normalized wavenumber $\kappa = k/k_1$ and the interval between the two transitions is given by the ratio $k_2/k_1 = 30$. 
    We fix the ratios between the potential slopes $A_1/A_2 = A_2/A_3 = 10$, and set $H^3e^{\alpha}/A_1$ such that $\mathcal{P}_{\mathcal{R}}\to2\times 10^{-9}$ for $\kappa\to0$. 
    The plots show different power spectra obtained by varying the value of the $\alpha$ parameter in the initial $\alpha$-vacuum state.}
    \label{fig: case:a1}
\end{figure}

\begin{figure}
    \centering
    \begin{minipage}[b]{0.49\textwidth}
        \centering
        \includegraphics[width=\textwidth]{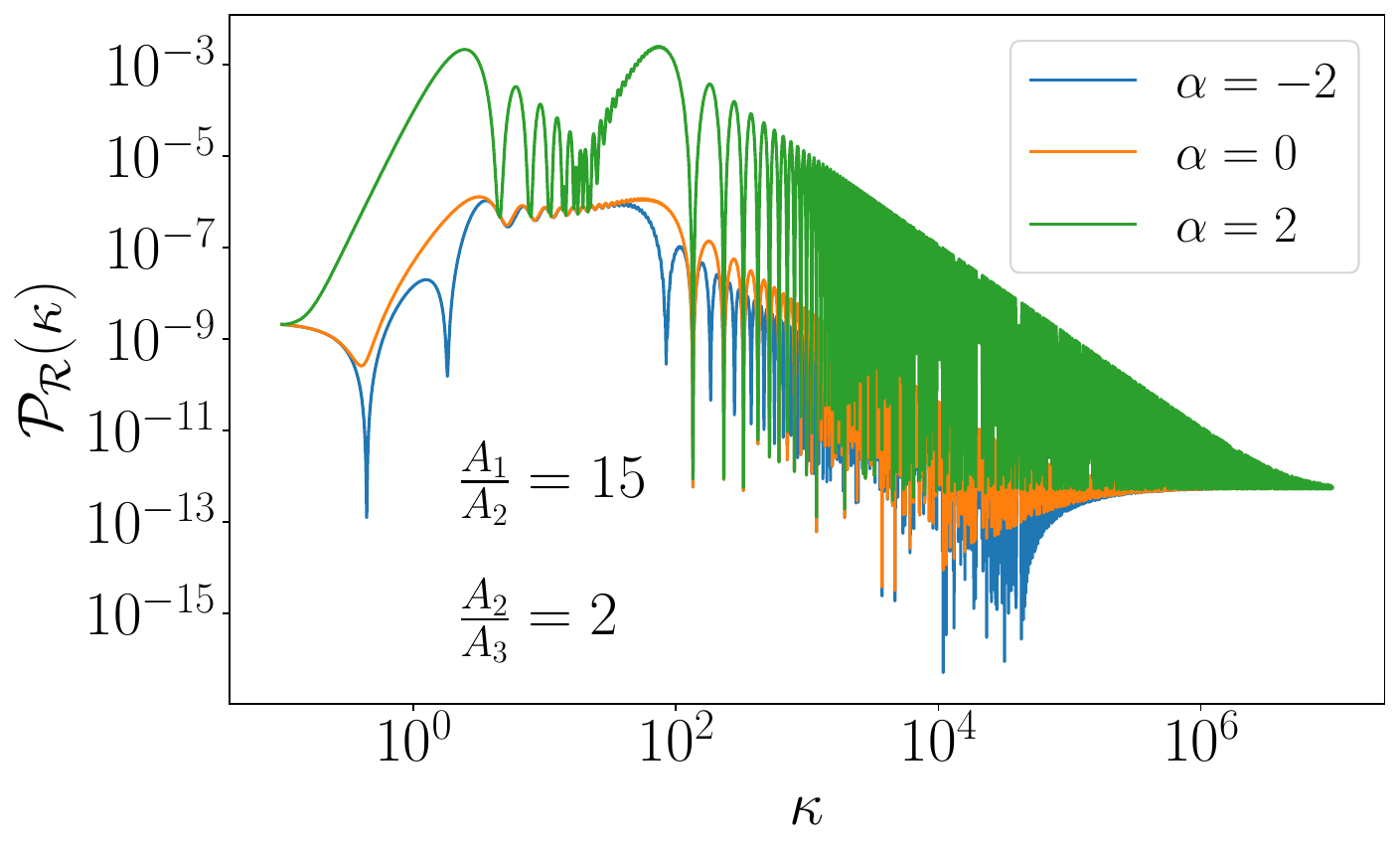}
    \end{minipage}
    \begin{minipage}[b]{0.49\textwidth}
        \centering
        \includegraphics[width=\textwidth]{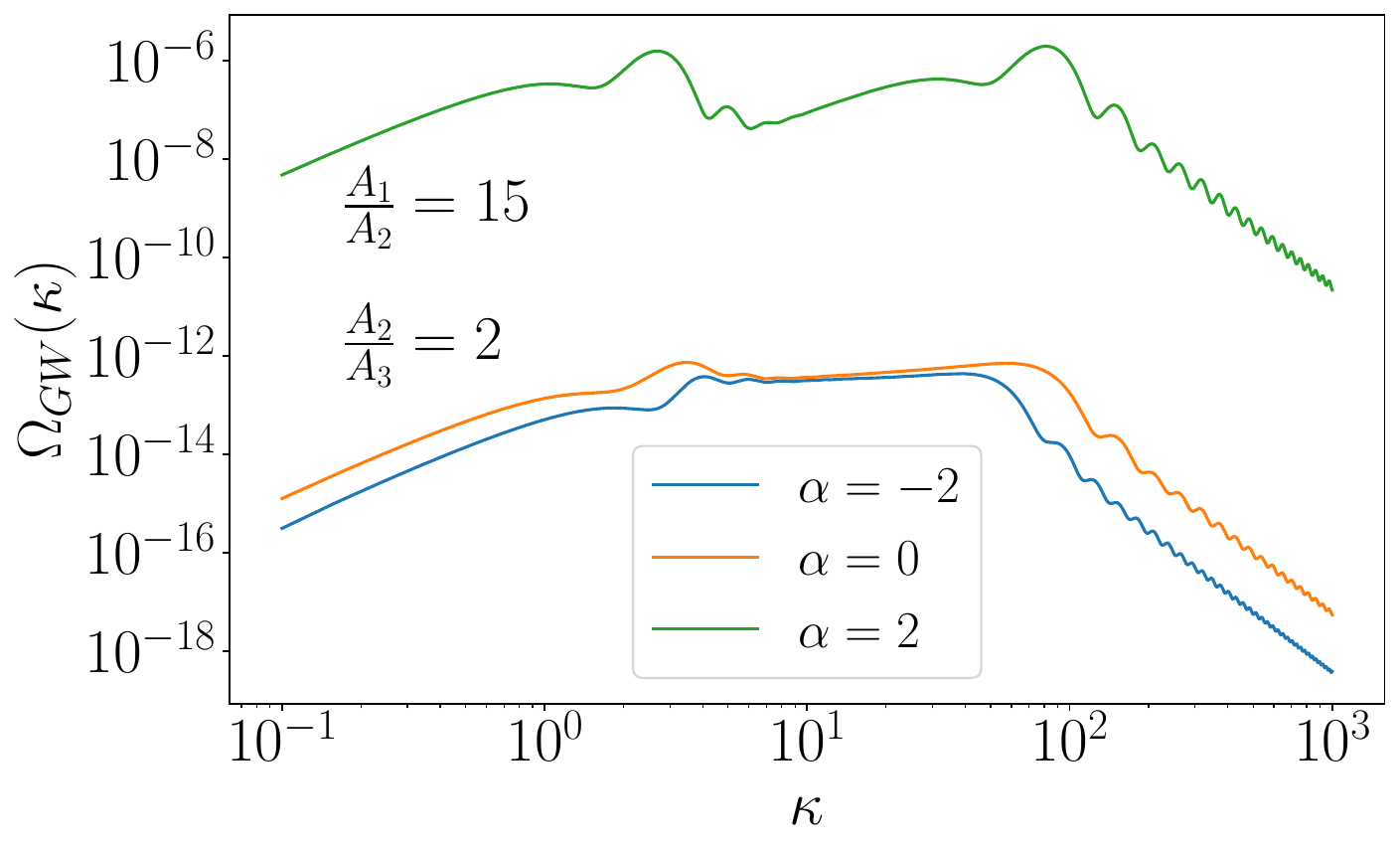}
    \end{minipage}
    \caption{\textit{Left}: The primordial power spectrum. \textit{Right}: The gravitational waves energy density for the Scalar Induced Gravitational Waves $\Omega_{GW}$. 
    Both the functions are shown as a function of the normalized wavenumber $\kappa = k/k_1$ (where $k_1$ is the wavenumber which crosses the horizon at the first transition) and the interval between the two transitions is given by the ratio $k_2/k_1 = 30$. 
    In this case, we are setting the ratios between the regimes $\frac{A_1}{A_2} = 15$ and $ \frac{A_2}{A_3} = 2$.
    The plot shows different behaviour by varying the value of the $\alpha$ parameter.}
    \label{fig: case:a2}
\end{figure}

\begin{figure}
    \centering
    \begin{minipage}[b]{0.49\textwidth}
        \centering
        \includegraphics[width=\textwidth]{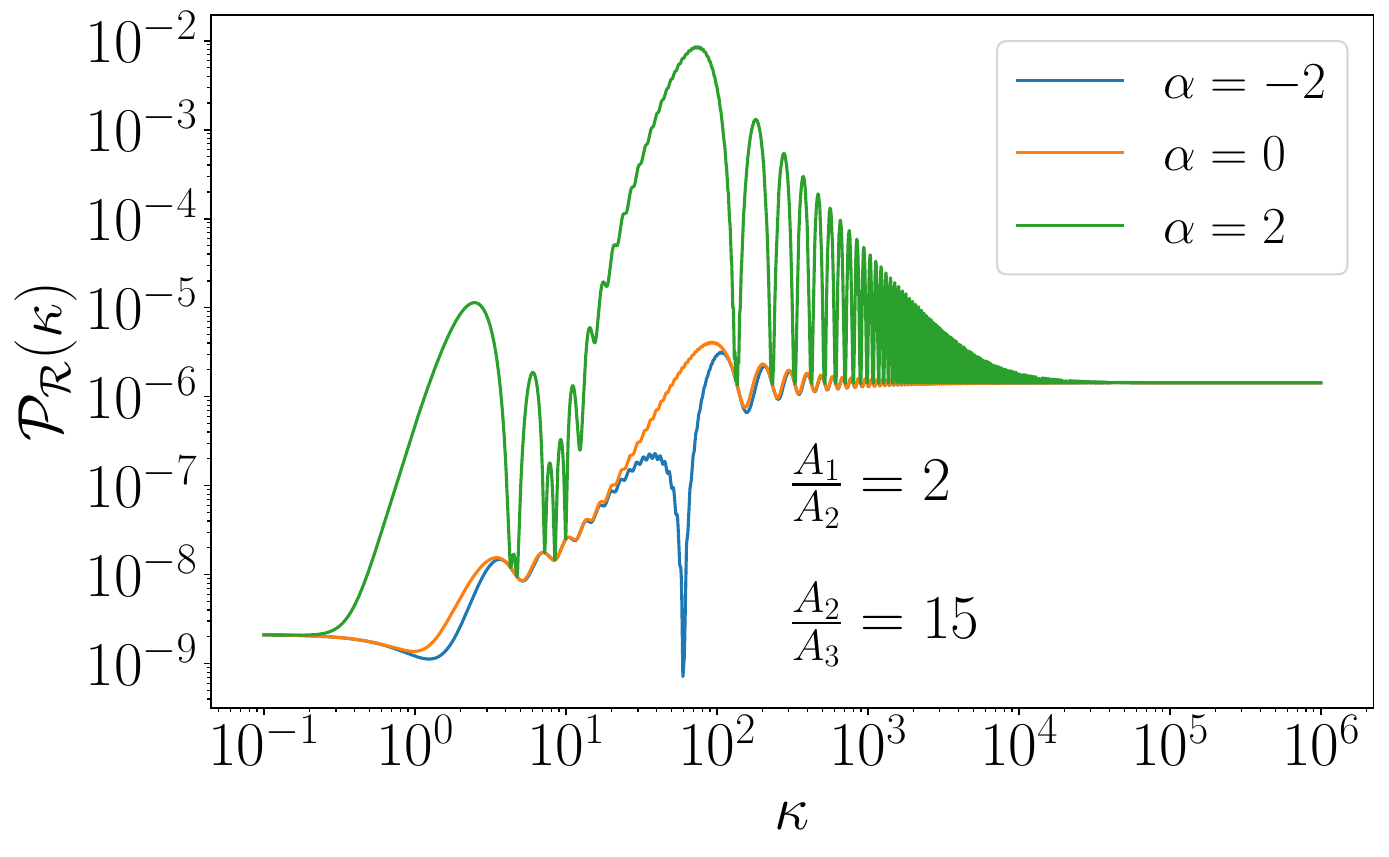}
    \end{minipage}
    \begin{minipage}[b]{0.49\textwidth}
        \centering
        \includegraphics[width=\textwidth]{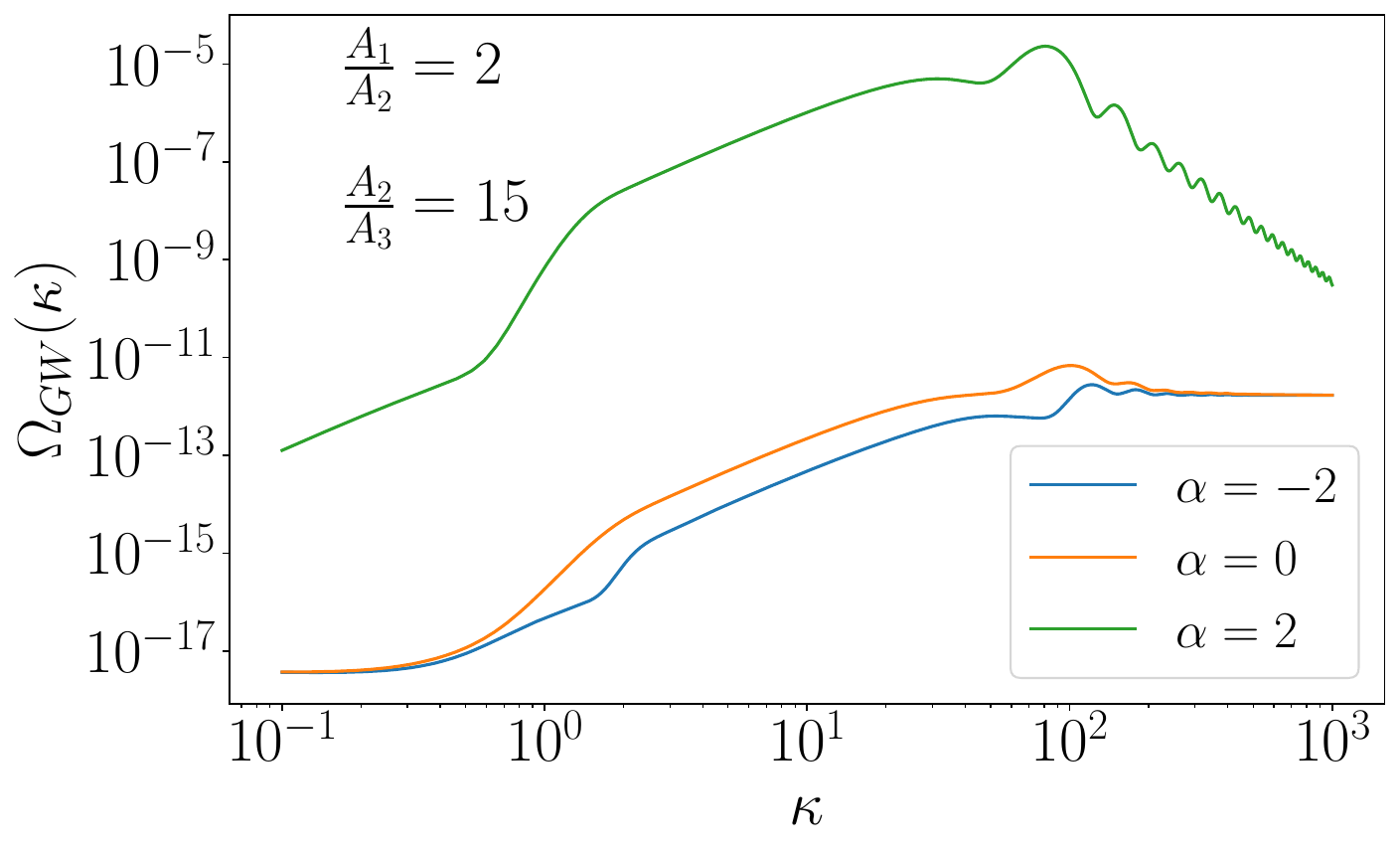}
    \end{minipage}
    \caption{\textit{Left}: The primordial power spectrum. \textit{Right}: The gravitational waves energy density for the Scalar Induced Gravitational Waves $\Omega_{GW}$. 
    Both the functions are plotted as a function of the normalized wavenumber $\kappa = k/k_{T_1}$ at which the first transition happens. In this case, we are setting the ratios between the regimes $\frac{A_1}{A_2} = 2$ and$ \frac{A_2}{A_3} = 15$. The ratio of the wavenumbers corresponding to the two transitions is set to $k_2/k_1 = 30$. The plot shows different behaviour by varying the value of the $\alpha$ parameter.}
    \label{fig: case:a3}
\end{figure}

\begin{figure}
    \centering
    \begin{minipage}[b]{0.48\textwidth}
        \centering
        \includegraphics[width=\textwidth]{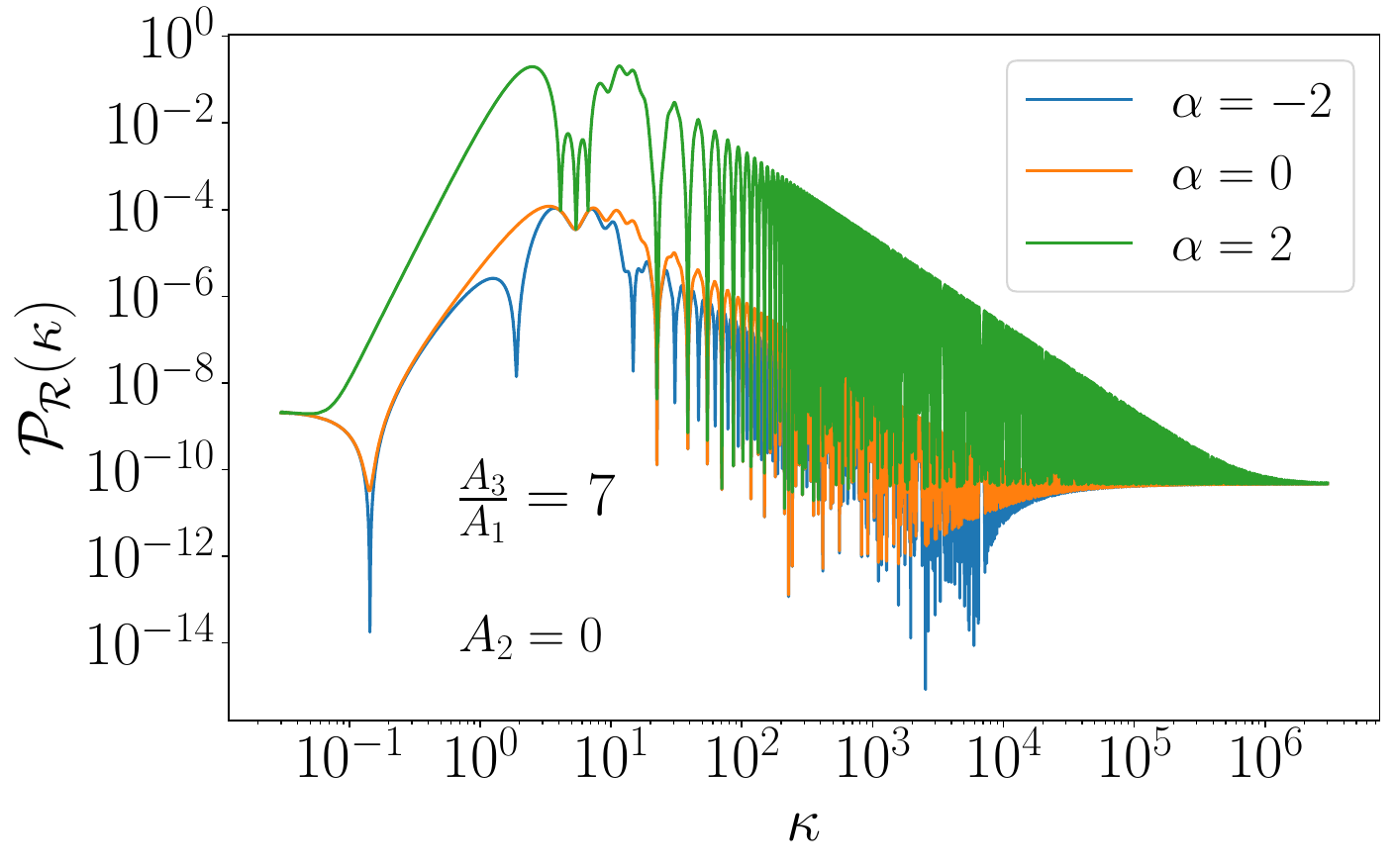}
    \end{minipage}
    \begin{minipage}[b]{0.49\textwidth}
        \centering
        \includegraphics[width=\textwidth]{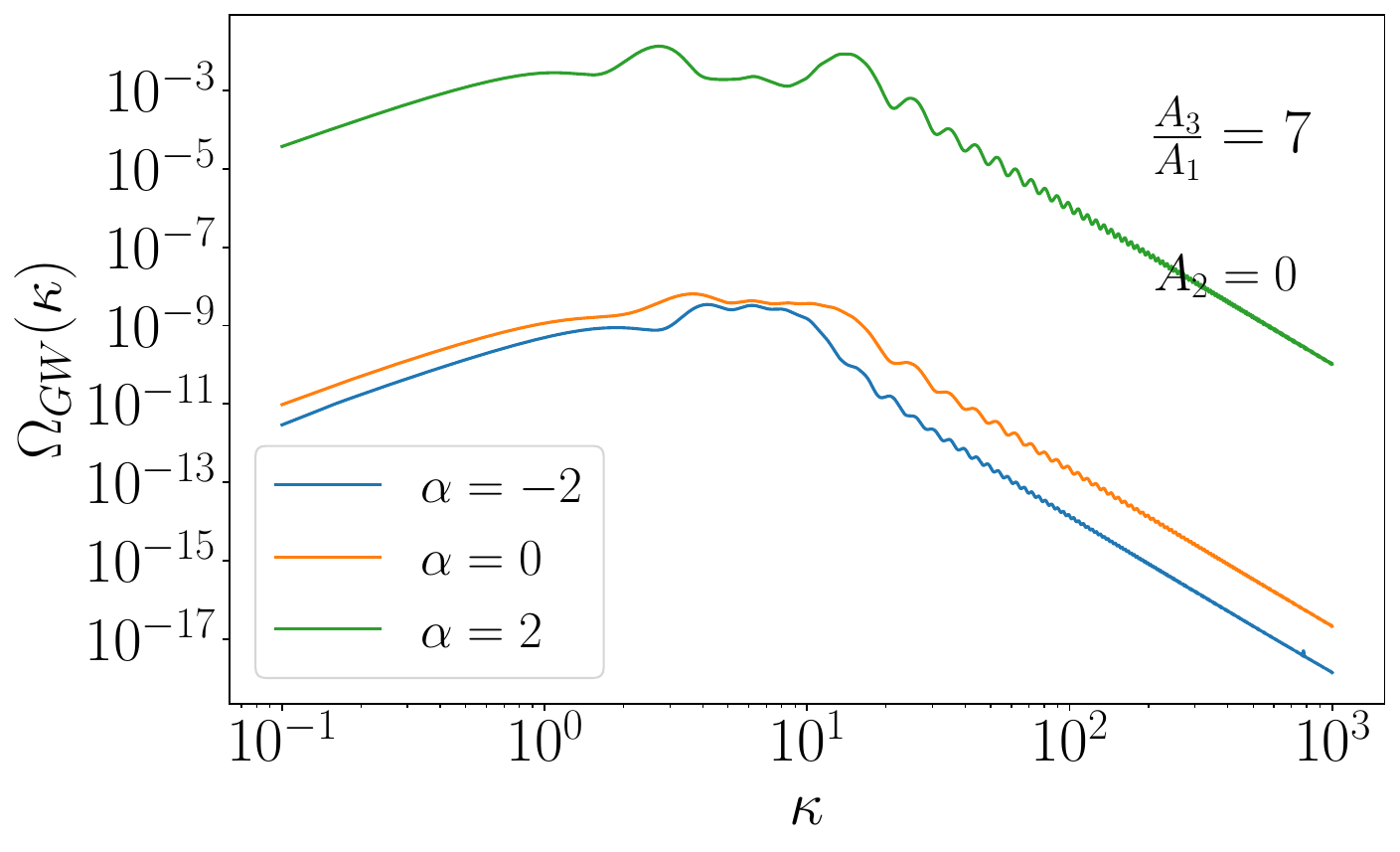}
    \end{minipage}
    \caption{\textit{Left}: The primordial power spectrum. \textit{Right}: The gravitational waves energy density for the Scalar Induced Gravitational Waves $\Omega_{GW}$. 
    Both the functions are plotted as a function of the normalized wavenumber $\kappa = k/k_{1}$ at which the first transition happens. In this case, we are setting the ratios between the regimes $\frac{A_3}{A_1} = 7 $ while keeping $A_2 = 0$. The ratio of the wavenumbers corresponding to the two transitions is set to $k_2/k_1 = 5$. The plot shows different behaviour by varying the value of the $\alpha$ parameter.}
    \label{fig: case:b1}
\end{figure}

In this section, we apply the previous formulae for multiple transitions to the specific case of $n = 2$ transitions in the Starobinsky piecewise-linear model \eqref{multipiecewisepotential}. 
In this case, there are two values of the inflation field, $\phi_1$ and $\phi_2$, at which the slope of the potential, $V(\phi)$, changes abruptly. 
The dynamics therefore can be split into three regimes. We assume that $A_1>0$ and the field starts in the slow roll attractor \eqref{eq:SRattractor1} for $\phi>\phi_1$, while for $\phi_2<\phi<\phi_1$ or $\phi<\phi_2$ the field velocity is given by Eq.~\eqref{dotphi-sum}. 
The general solution for the mode function is given by Eqs.~\eqref{eq. general solution} and \eqref{freefk}, with arbitrary constants of integration, ${\mathcal{A}}_{k,i}$ and ${\mathcal{B}}_{k,i}$, to be determined. We start with a de Sitter $\alpha$-vacuum state (corresponding to $\beta=0$ in \eqref{alpha-vacuum}) and at each transition, we apply the Bogoliubov matching \eqref{A+Brecursive}.
In our numerical examples, we explore the effect on the power spectra of the choice of $\alpha$ parameter for the initial quantum state and the two ratios between the different slopes in the three regimes. We fix the interval between the two transitions 
such that first $k_2 = 30 k_1$, so the transitions are relatively well separated, but later we fix $k_2 = 5 k_1$ (such that the scalar field reaches the second transition at $\phi_2$ when $A_2<0$). 
%Let's then show the analytical results for the Bogoliubov coefficients in this case and then we can move to analyzing the numerical results we obtain for the primordial power spectrum and the spectrum for the secondary gravitational waves. 
%\MCcomment{to be honest... the analytical expression for the Bogoliubov coefficients, in this case, is quite lengthy... it would take two pages... I don't know if we should add it.}
We explore the primordial power spectrum~\eqref{PRlate} and resulting stochastic gravitational wave spectrum~\eqref{OmegaGWeq} in the different regimes: 
\begin{enumerate}
    \item[a)] In Figures~\ref{fig: case:a1}, \ref{fig: case:a2}, \ref{fig: case:a3} we vary $\alpha$ for different ratios between the potential gradients while keeping the slopes positive throughout ($A_i>0$). In each case, we see two distinct peaks, with steeper growth and much stronger enhancement, for $\alpha>0$, as we saw for a single transition, whereas for $\alpha\leq0$ the first peak height is lower and the peak is much less clearly defined.
    The scalar power spectra in Figure~\ref{fig: case:a1}, where $A_2/A_3=A_1/A_2$, and Figure~\ref{fig: case:a3}, where $A_2/A_3\gg A_1/A_2$, are similar, both showing two main peaks for $\alpha>0$, with the second peak at $k/k_2\simeq \pi$ (which corresponds to $\kappa\simeq 30\pi$) higher than the first peak at $\kappa=k/k_1\simeq \pi$. Meanwhile in Figure~\ref{fig: case:a2}, where $A_2/A_3\ll A_1/A_2$, the first peak is approximately the same height as the second peak for any $\alpha$. As a result $\Omega_{GW}$ steadily increases to a single peak at $\kappa\simeq 30\pi$ in Figures~\ref{fig: case:a1} and \ref{fig: case:a3}, while in Figure~\ref{fig: case:a2} the GW power spectrum has two approximately equal-height peaks for $\alpha>0$,  or an extended plateau for $\alpha\leq0$. 
    %$\kappa\simeq 3$ as well as $\kappa\simeq 10^2$. 
    \item[b)] In Figures~\ref{fig: case:b1}, \ref{fig: case:b2} and \ref{fig: case:b3}, we explore the behaviour of the power spectra in cases where the inflationary potential is completely flat ($A_2=0$) in the intermediate region $\phi_2<\phi<\phi_1$, while allowing for different ratios for the initial and final slopes ($A_1/A_3$). Note that in this case for the field to evolve beyond $\phi_2$ we require $\phi_1-\phi_2<A_1/9H^2$, otherwise the field comes to a stop before reaching $\phi_2$.
    %($\phi\to\phi_\infty>\phi_2$).
    %
    In all of the configurations that we investigate in Figures~\ref{fig: case:b1}, \ref{fig: case:b2} and \ref{fig: case:b3} we find very similar results for the scalar power spectrum and for the GW power spectrum. 
    As might be expected, the first peak height is more strongly enhanced in all cases compared with the equivalent case where $A_2>0$. As before, the two peaks, corresponding to the two transitions, are much more clearly defined when $\alpha>0$ and $\Omega_{GW}$ is characterized by a double peak, where the second peak is slightly lower than the first one. 
    %\MCcomment{note that the amplitudes for both the PS and Omega would suggest that in this case, a higher value for $\alpha$ would be allowed}
    \item[c)] Lastly, we explore in Figure~\ref{fig: case:c} a configuration where the slope in the intermediate regime, $A_2$, takes a negative value. As for $A_2=0$, we must ensure that $\phi_1-\phi_2$ is small enough (or the transient velocity for $\phi<\phi_1$ is large enough) that the field does indeed reach $\phi_2$. Otherwise, the field stops ($\dot\phi=0$) and it starts evolving back towards $\phi_1$, where it becomes trapped.
    This implies that for a given value of $A_2<0$ the ratio $k_2/k_1$ is bounded from above; we require:
    \begin{equation}
        1<\frac{k_2}{k_1} < \left( 1 - \frac{A_1}{A_2} \right)^{1/3} \,.
    \end{equation}
    For all values of $\alpha$ we see a significant enhancement of the scalar power spectrum at the scale corresponding to the first transition ($\kappa=1$). When $\alpha>0$ there is a much steeper, greater enhancement and the power spectrum clearly drops before the second peak. This occurs even though we have set $A_3=A_1$ in this example, so that there is no overall change in slope between early and late times, with only a transient flat regime. Nonetheless, for any value of $\alpha$ there is a strong effect on both the scalar and induced GW power spectra across a wide range of scales, e.g., $10^{-1}<\kappa<10^4$ for $\alpha=2$.    
\end{enumerate}  
In all of the three cases discussed above, we can see that we obtain a distinctive structure of peaks in the power spectra corresponding to the frequencies at which the two transitions occur, with a positive value for $\alpha$ distinguished by steeper growth, stronger enhancement, and two distinct peaks. In all cases, the scalar and tensor power spectra approach the same values in the low and high $\kappa$ limits regardless of the value of $\alpha$. That is because we fix the normalisation at long wavelengths before the transition, while the asymptotic values at short wavelengths are then determined solely by the ratio of the initial and final slopes, $A_3/A_1$.

{\begin{figure}
    \centering
    \begin{minipage}[b]{0.49\textwidth}
        \centering
        \includegraphics[width=\textwidth]{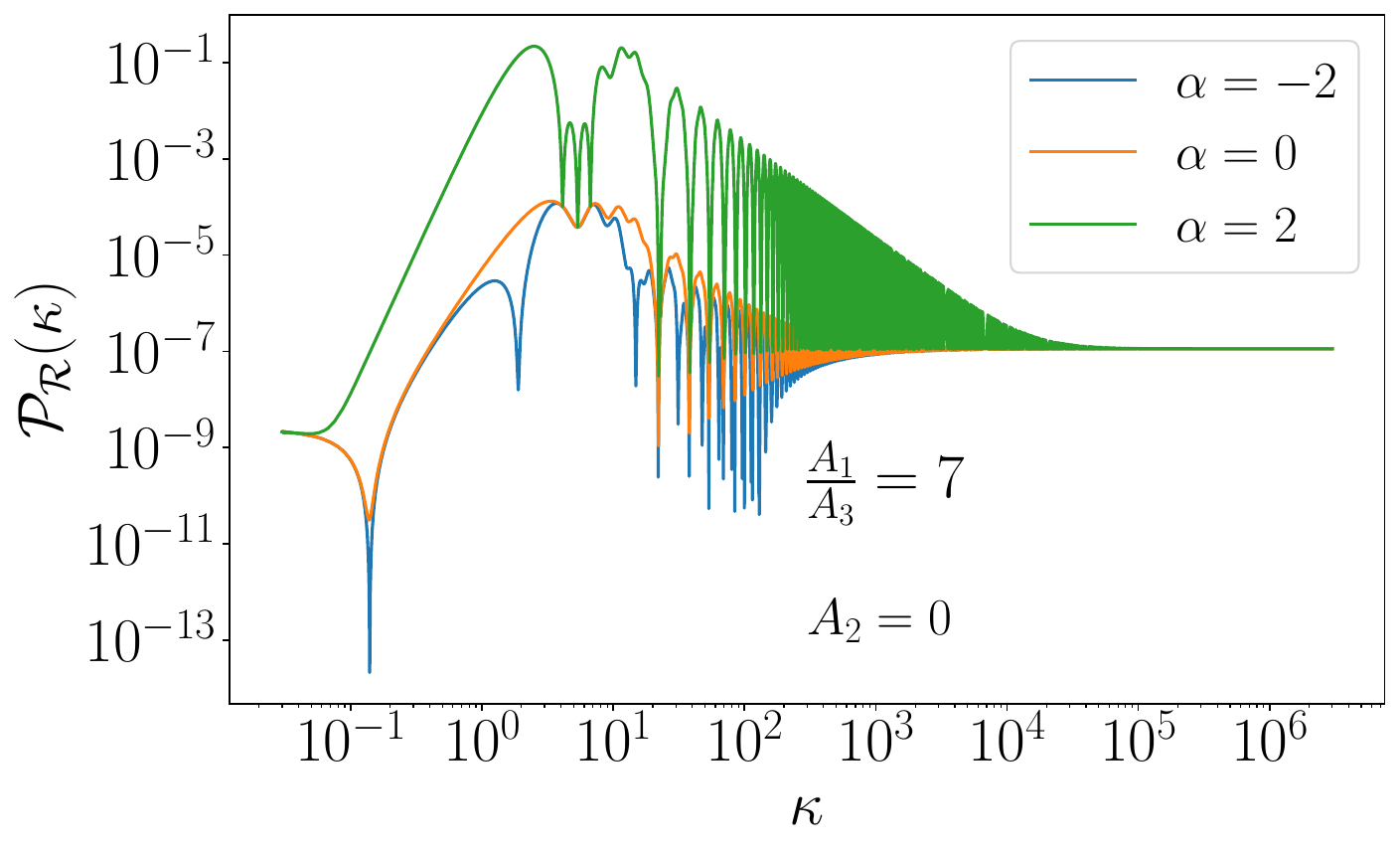}
    \end{minipage}
    \begin{minipage}[b]{0.49\textwidth}
        \centering
        \includegraphics[width=\textwidth]{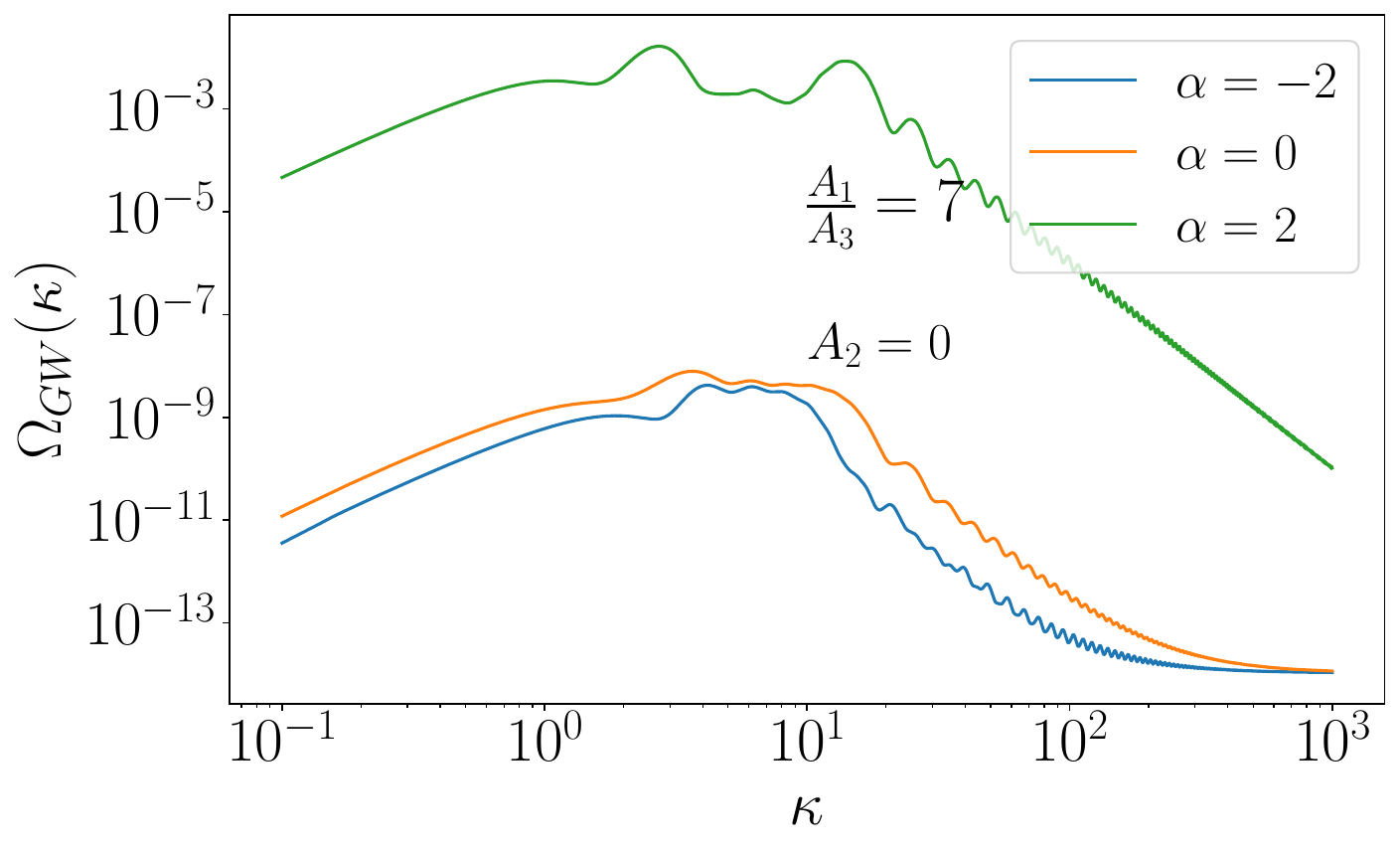}
    \end{minipage}
    \caption{\textit{Left}: The primordial power spectrum. \textit{Right}: The gravitational waves energy density for the Scalar Induced Gravitational Waves $\Omega_{GW}$. 
    Both the functions are plotted as a function of the normalized wavenumber $\kappa = k/k_{T_1}$ at which the first transition happens. In this case, we are setting the ratios between the regimes $A_3/A_1 = 1/7$ while keeping $A_2 = 0$. The ratio of the wavenumbers corresponding to the two transitions is set to $k_2/k_1 = 5$. The plot shows different behaviour by varying the value of the $\alpha$ parameter.}
    \label{fig: case:b2}
\end{figure}

\begin{figure}
    \centering
    \begin{minipage}[b]{0.49\textwidth}
        \centering
        \includegraphics[width=\textwidth]{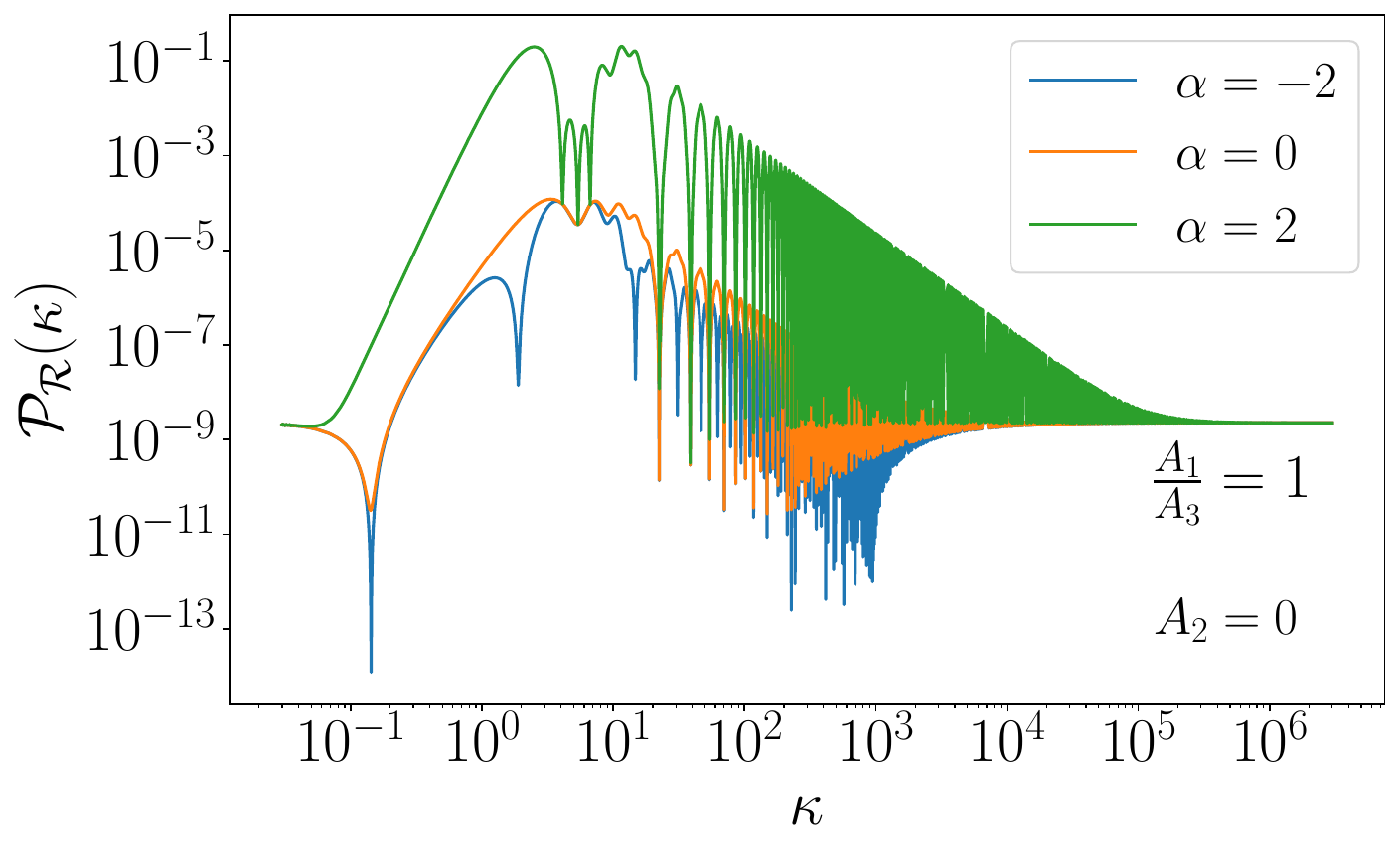}
    \end{minipage}
    \begin{minipage}[b]{0.5\textwidth}
        \centering
        \includegraphics[width=\textwidth]{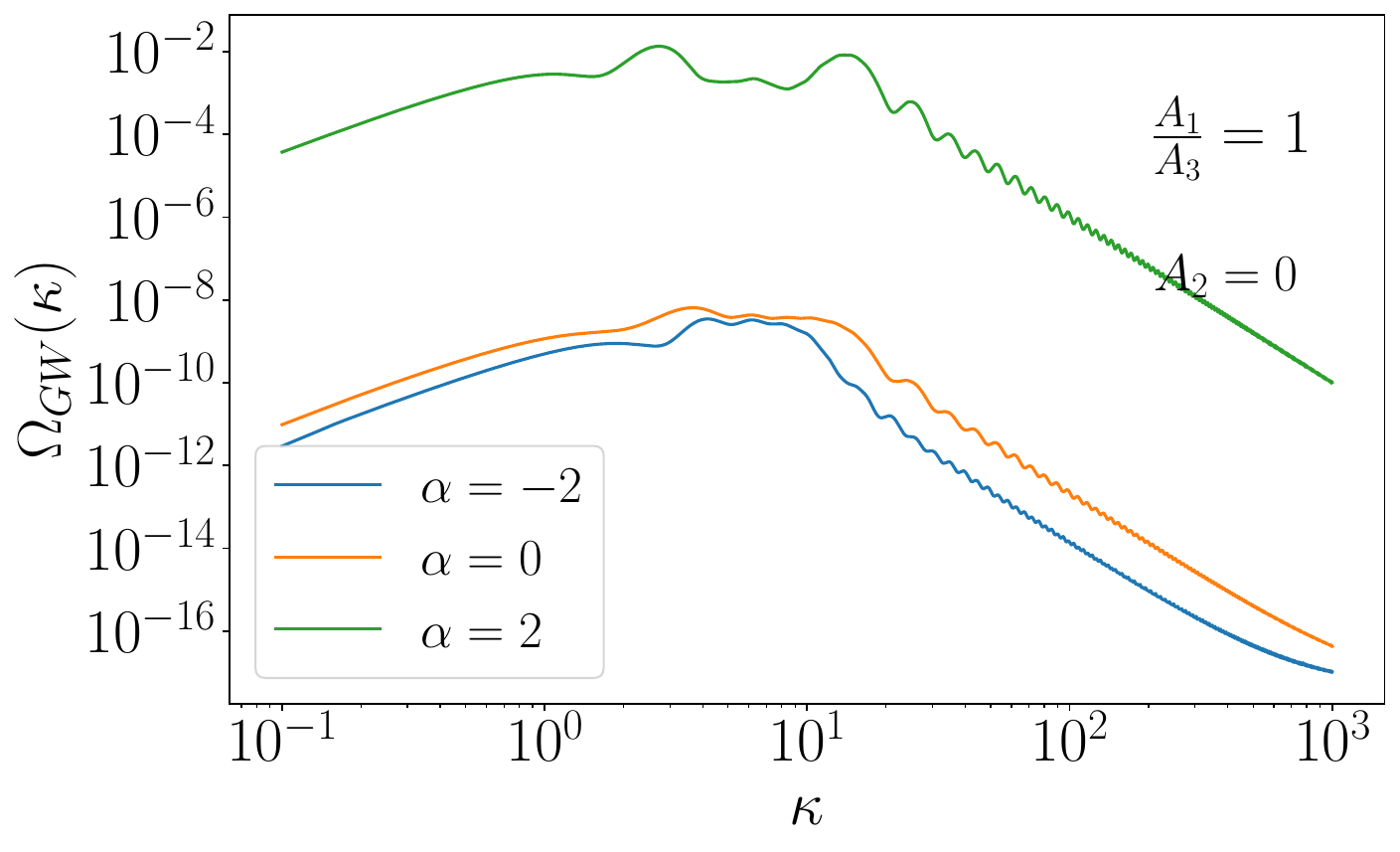}
    \end{minipage}
    \caption{\textit{Left}: The primordial power spectrum. \textit{Right}: The gravitational waves energy density for the Scalar Induced Gravitational Waves $\Omega_{GW}$. 
    Both the functions are plotted as a function of the normalized wavenumber $\kappa = k/k_{T_1}$ at which the first transition happens. In this case, we are setting the ratios between the regimes $\frac{A_1}{A_3} = 1 $ while keeping $A_2 = 0$. The ratio of the wavenumbers corresponding to the two transitions is set to $k_2/k_1 = 5$. The plot shows different behaviour by varying the value of the $\alpha$ parameter.}
    \label{fig: case:b3}
\end{figure}

\begin{figure}
    \centering
    \begin{minipage}[b]{0.49\textwidth}
        \centering
        \includegraphics[width=\textwidth]{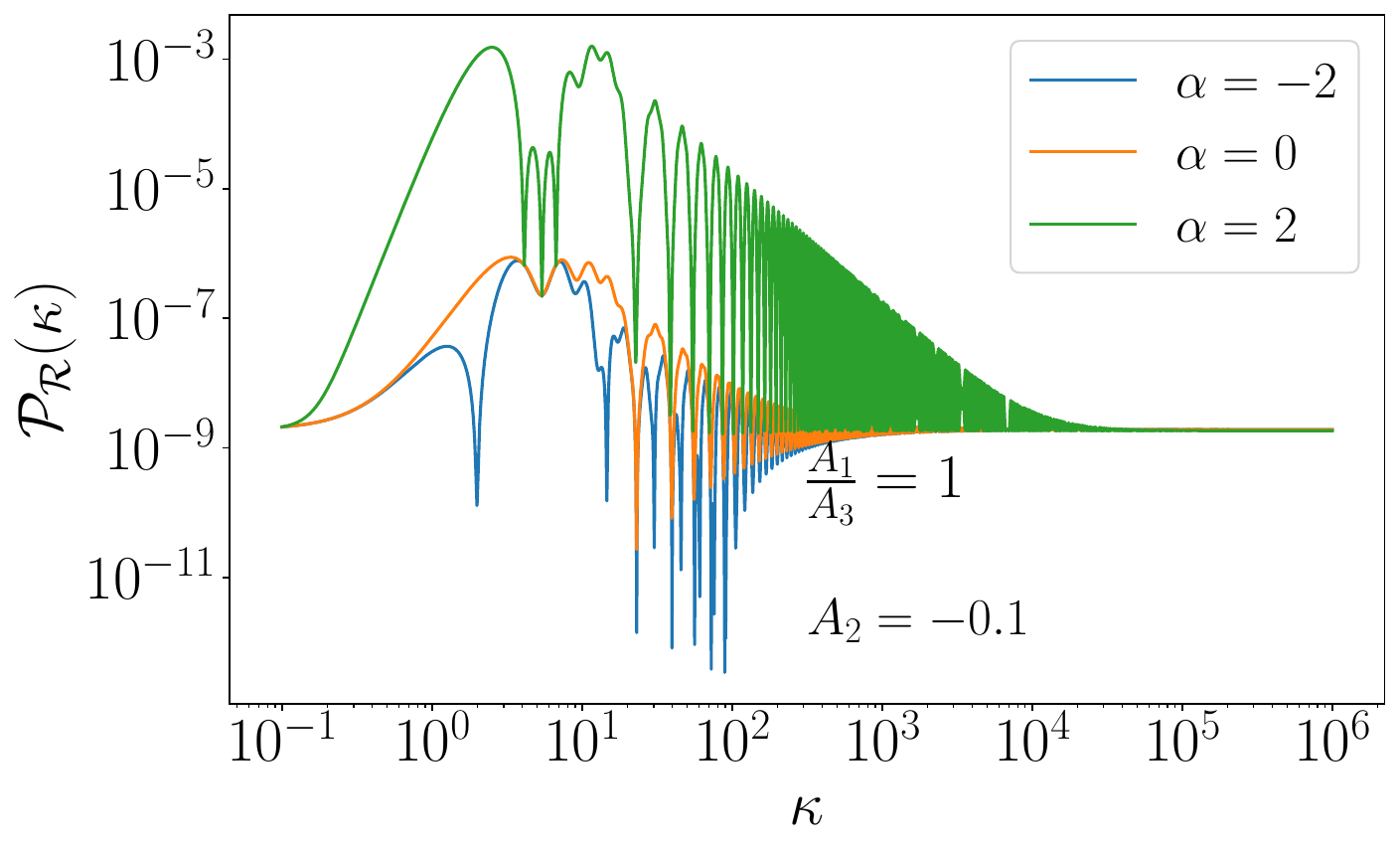}
    \end{minipage}
    \begin{minipage}[b]{0.49\textwidth}
        \centering
        \includegraphics[width=\textwidth]{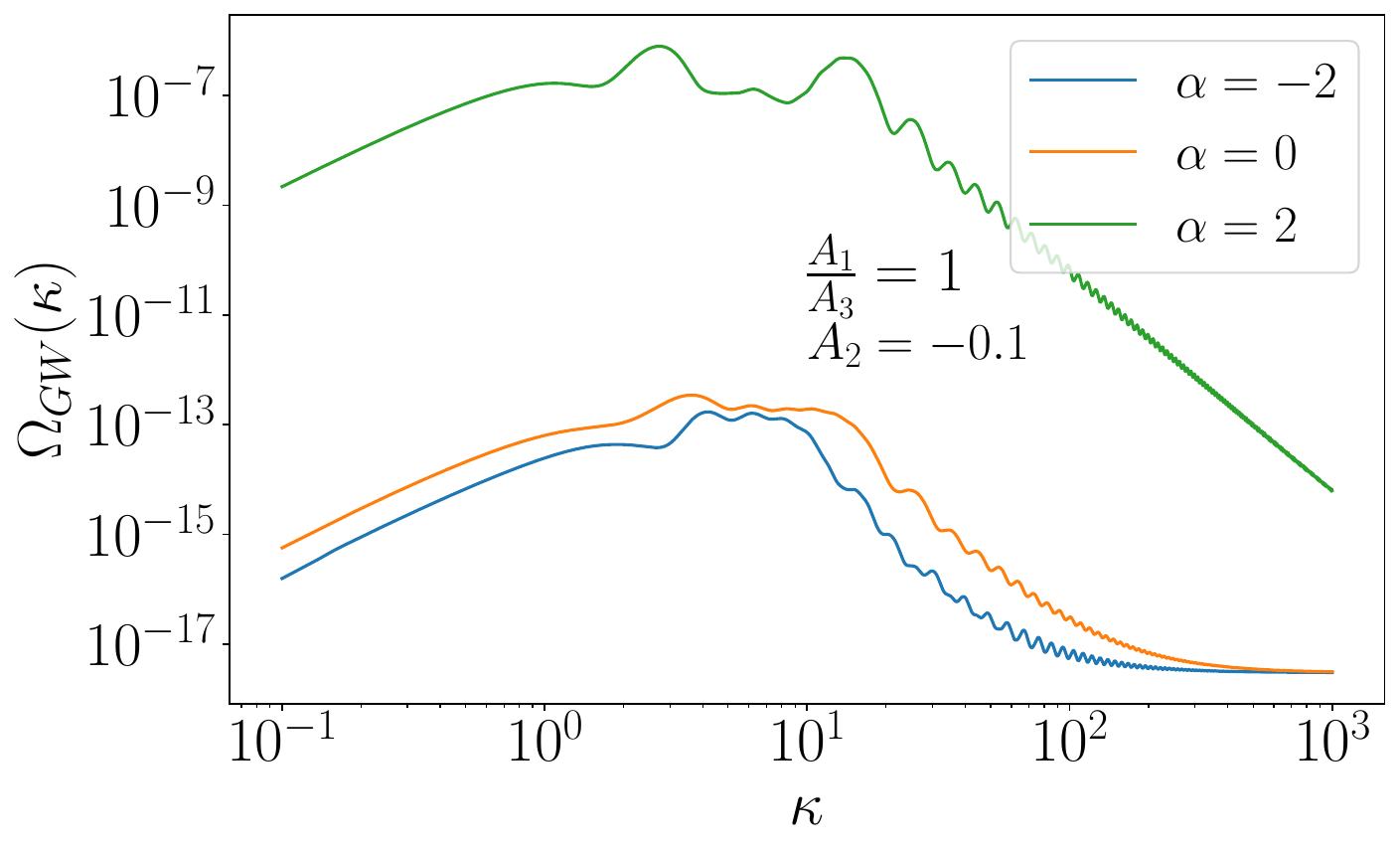}
    \end{minipage}
    \caption{\textit{Left}: The primordial power spectrum. \textit{Right}: The gravitational waves energy density for the Scalar Induced Gravitational Waves $\Omega_{GW}$. 
    Both the functions are plotted as a function of the normalized wavenumber $\kappa = k/k_{T_1}$ at which the first transition happens. In this case, we are setting the ratios between the regimes $\frac{A_1}{A_3} = 1 $ while considering a negative value for $A_2 = -0.1$, consistent with the constraint. The ratio of the wavenumbers corresponding to the two transitions is set to $k_2/k_1 = 5$. The plot shows different behaviour by varying the value of the $\alpha$ parameter.}
    \label{fig: case:c}
\end{figure}

\section{Conclusions}
\label{sec:conclusions}

Models featuring sudden transitions in the scalar potential for curvature perturbations have attracted significant attention in recent years due to their key role in the possible production of primordial black holes and an associated scalar-induced gravitational wave background. Two key factors in determining the abundance and mass spectrum of PBHs as well as the spectrum of induced GWs are the height of the peaks in the power spectrum and the steepness of its growth. 
In this work, we have extended previous studies by considering a scenario where the quantum field originates in an excited state, specifically a de Sitter-invariant $\alpha$-vacuum~\cite{Allen:1985ux}. Furthermore, we have studied a generalization of Starobinsky’s piecewise-linear potential to include an arbitrary number of sudden transitions, providing an exact, order-by-order prescription for computing the Bogoliubov coefficients. These coefficients are sufficient to determine the primordial scalar power spectrum at the end of inflation.
Our analytical results for a single sudden transition demonstrate a steep growth of the scalar power spectrum proportional to $k^6$ for $\alpha$-vacua with $\alpha>0$, in contrast to the expected growth proportional to $k^4$ for an initial Bunch-Davies vacuum state where the $\alpha$ parameter is set to zero.
Since the scalar perturbations source primordial gravitational waves at second order, we have numerically calculated the resulting gravitational wave energy density. Our findings are consistent with the previous results in the literature for an initial Bunch-Davies vacuum while revealing a steeply peaked GW power spectrum for positive values of $\alpha$. These results are particularly interesting in light of the recent PTA results \cite{NANOGrav:2023gor, NANOGrav:2023hvm} and possible observations with LISA \cite{LISACosmologyWorkingGroup:2022jok} and other future GW detectors.
%\DWcomment{I'm not sure about the relevance of current and future CMB observations which are very large scale: \cite{Bartolo:2016ami, BICEP2:2018kqh, BICEP:2021xfz, CMB-S4:2016ple, Abazajian:2019tiv, LiteBIRD:2022cnt, Campeti:2020xwn, Planck:2018vyg, Punturo:2010zz, Kawamura:2006up, Crowder:2005nr}.}
For the case of two sudden transitions, we explored different possible combinations of the three slopes characterizing the generalized piecewise-linear potential. From the perspective of the curvature power spectrum, we demonstrated the possibility of generating multiple peaks with varying heights. For initial $\alpha$-vacua with $\alpha>0$ we find clearly separated peaks, whereas for $\alpha\leq0$ the initial peaks are less pronounced. Multiple peaks in the scalar power spectrum could lead to a non-trivial mass spectrum for PBHs which itself could lead to an interesting phenomenology. 
In all cases we have calculated the corresponding scalar-induced gravitational wave background, resulting in distinctive signal templates that could be tested by future experiments. These experiments may provide valuable insights into the initial quantum state of the universe.
Our results suggest that localised features in the primordial scalar perturbations produced during sudden transitions during inflation could reveal information about both the nature of the transition and the incoming quantum state. Of course, it is unclear at this stage whether information about the initial state could be extracted from observations of gravitational relics such as PBHs and GWs independently of knowledge about the transition. We have only investigated instantaneous models of the transition in this paper, and we know this will give spurious effects on arbitrarily small scales. Any finite duration of the transition, $\Delta t$,  is expected to smooth out structures on a scale $k\gg a/\Delta t$. 
The resulting abundance of PBHs can also be very sensitive to non-Gaussianity in the density distribution~\cite{Bullock:1996at,Young:2013oia,Yoo:2019pma,Biagetti:2021eep,Kitajima:2021fpq,Ferrante:2022mui,Gow:2022jfb,Matsubara:2022nbr}, which arise from higher-order corrections to the linear analysis of perturbations used in this paper. Indeed non-perturbative techniques are ultimately needed to study very large, very rare perturbations which lead to PBHs~\cite{Pattison:2017mbe,Biagetti:2018pjj,Ezquiaga:2019ftu,Pi:2022ysn,Cai:2022erk}. It would be interesting to extend our study of non-Bunch-Davies initial quantum states to such non-linear analyses. 
%    Two-transition configurations where the slope in the intermediate regime, $A_2$, takes a negative value exhibits an even more interesting behaviour, even when we set $\alpha = 0$, since it differs slightly from the findings in \cite{Pi:2022zxs}. 
%     \MCcomment{actually this is a feature shared with also other cases. In Shi Pi's paper the SIGW when they reach the peak freeze basically to a value close to such peak, while in most of our cases, we reach a peak and then they drop down and take a lower asymptotic value}

\acknowledgments
M.~C.~is grateful to the Insitute of Cosmology and Gravitation for hospitality and support. The work of M.~C.~is supported by the ``Angelo Della Riccia'' Fellowship. The work of M.~C., G.~M., and O.~P.~is partially supported by the INFN “Theoretical Astroparticle Physics” (TAsP) project. O.~P.~and G.~M.~acknowledge support by Ministero dell’Universit`a e della Ricerca (MUR), PRIN2022 program (Grant PANTHEON 2022E2J4RK) Italy. 
D.W.~was supported by the Science and Technology Facilities Council (grant number ST/W001225/1).
For the purpose of open access, the authors have applied a Creative Commons Attribution (CC-BY) licence to any Author Accepted Manuscript version arising from this work.

\bibliographystyle{JHEP}
\bibliography{ref}

\end{document}